\documentclass[aps,prb,twocolumn,showpacs,superscriptaddress]{revtex4-1}
\usepackage{bm, amsmath,amssymb}
\usepackage{textcomp}
\usepackage{gensymb}
\usepackage{graphicx}
\usepackage{graphics}
\usepackage{longtable}
\usepackage{xcolor}

\begin{document}

\title{Surfactant mediated growth of ferromagnetic Mn $\delta$-doped Si}

\author{S. Kahwaji}
\email{sam@dal.ca}
\affiliation{Department of Physics and Atmospheric Science, Dalhousie University, Halifax, Nova Scotia, Canada B3H 3J5}
\author{R.A. Gordon}
\affiliation{PNCSRF/CLS, APS Sector 20, Argonne, IL, USA 60439}
\author{E.D. Crozier}
\affiliation{Department of Physics, Simon Fraser University, Burnaby, British Columbia, Canada V5A 1S6}
\author{S. Roorda}
\affiliation{D\'{e}partement de Physique, Universit\'{e} de Montr\'{e}al, Montr\'{e}al, Qu\'{e}bec, Canada, H3C 3J7}
\author{M. D. Robertson}
\affiliation{Department of Physics, Acadia University, Wolfville, Nova Scotia, Canada B4P 2R6}
\author{J. Zhu}
\affiliation {National Renewable Energy Laboratory, Golden, CO 80401, USA.}
\author{T.L. Monchesky}
\affiliation{Department of Physics and Atmospheric Science, Dalhousie University, Halifax, Nova Scotia, Canada B3H 3J5}

\date{\today}

\begin{abstract}
We present an investigation of Mn $\delta$-doped layers in Si(001) grown by molecular beam epitaxy.  We discovered that a Pb surfactant has significant effect on the structural and magnetic properties of the submonolayer of Mn, which depends on the Si capping layer growth temperature, $T_{\textrm{Si}}$, and the Mn coverage, $\theta_{\textrm{Mn}}$. The results presented in this paper identify three regions in the growth-phase-diagram characterized by distinct magnetic behaviors and crystal structures. In one region, X-ray absorption fine structure (XAFS)  and transmission electron microscopy (TEM) experiments indicate that MnSi nanocrystallites form with B2-like crystal type structures. At the optimal growth conditions, $T_{\textrm{Si}}$ = 200~\celsius~and $\theta_{\textrm{Mn}} = 0.26$ monolayer, a ferromagnetic phase develops with a Curie temperature $T_C >400$~K and a saturation moment $m_{sat}$= 1.56~$\mu_B$/Mn, whereas $T_C$ drops to zero for a control sample prepared without Pb. For $T_{\textrm{Si}} > 200$~\celsius, MnSi-B20 type precipitates form with a $T_C \approx 170$~K.  Rutherford backscattering spectroscopy shows that the increase in the remanent magnetization in these two phases is possibly correlated with an increase in the Mn substitutional fraction, which suggests that a Si$_{1-x}$Mn$_x$ dilute magnetic semiconductor may be forming in the matrix between the precipitates. Density functional calculations show that Pb changes the pathway by which the Mn atoms access the Si substitutional sites, Mn$_{\textrm{Si}}$.  While the Pb increases the formation energy of Mn$_{\textrm{Si}}$ at the Si surface, it enables substitutional incorporation by lowering the formation energy of Si vacancies by 0.92~eV.
\end{abstract}

\pacs{75.70.Ak, 61.05.cj, 75.50.Pp, 85.75.-d}

\keywords{spintronics, dilute magnetic semiconductors, ferromagnetic semiconductors, $\delta$-doped Si,MnSi}

\maketitle

\section{Introduction}

Dilute magnetic semiconductors (DMS) offer control over the carriers responsible for magnetic order, providing unique opportunities for modern electronics.\cite{Dietl:2010nm} Given the technological prevalence of Si, it is important to understand whether it is possible to create a Si-based DMS.  Mean field calculations predict relatively high Curie temperatures, $T_C$, for a dilute Si$_{1-x}$Mn$_{x}$  alloy,\cite{Dietl:2000sci} and density functional theory (DFT) calculations show that substitutional Mn in Si will produce a ferromagnetic semiconductor with a magnetic moment just below 3 $\mu_B$/Mn.\cite{Stroppa:2003prb} One of the challenges that Si presents for DMS is the high growth temperatures that are required for epitaxy.\cite{Eaglesham:1990prl}

Early attempts to create a homogeneous alloy by molecular beam epitaxy (MBE) demonstrate that the Mn phase segregates into nanocolumns.\cite{zhangJAP}  In Mn-implanted Si, post implantation annealing produces precipitates from the MnSi$_{1.7}$ family of phases.\cite{Affouda}  However, the nanocrystalline phases that result are interesting in their own right. Sputtered Mn$_{0.05}$Si$_{0.95}$ alloys \cite{Zhang} and Mn-implanted Si \cite{Bolduc:2005prb,Ko:2008jap} report Curie temperatures in excess of 400 K, in contrast to all known bulk Mn-Si phases that all have magnetic ordering temperatures below 47 K. The origin of this high-$T_C$ phase is not clear (see discussion in Ref. \onlinecite{Menshov:2011prb} and references therein). The magnetic properties that are sensitive to the preparation conditions,\cite{ZhouPRB75}have been suggested to be due to the presence of Mn impurities in the MnSi$_{1.7}$ phase.\cite{Orlov:2009jetp}  A recent proposal attributes the high $T_C$ to the interactions between the large magnetic moments of the impurities inside the precipitates.\cite{Menshov:2011prb}  The Si matrix must play a role in coupling these particles, although the exact mechanism is still not well understood.  A second interpretation suggests that carrier-mediated ferromagnetism due to Mn impurities in the Si matrix is responsible for the high $T_C$.\cite{Ko:2008jap}

Another way to create a DMS is by the deposition of submonolayers of magnetic impurities in a semiconductor host, referred to as $\delta$-doping, as first demonstrated in (Ga,Mn)As and (Ga,Mn)Sb.\cite{Kawakami:2000apl, Chen:2002apl}  DFT calculations show that a $\delta$-doped layer of substitutional Mn in Si is half-metallic.\cite{Qian:2006prl}   However, scanning tunneling microscopy studies show that interstitial sites are energetically more favorable,\cite{Krause:2007prb} which drives Mn to the subsurface where it forms wires.\cite{Liu:2008ss, Nolph:2011ss}  Interstitial sites are also interesting candidates for creating ferromagnetic order since DFT calculations show that a 1/4 $\delta$-doped layer of interstitial Mn in Si is half-metallic.\cite{Wu:2007prl}  A pure interstitial layer is likely difficult to achieve, due to the tendency of the Mn to form clusters,\cite{Bernardini:2004apl, Liu:2008prb} although a mixed $\delta$-doped layer may also yield interesting magnetic properties.\cite{Otrokov:2011prb}

In this paper, we investigate the influence of a Pb surfactant on the local structure and magnetism of $\delta$-doped Mn-layers on Si(001). In the case of Si, $\delta$-doping offers the advantage of independent control of the Mn and Si growth temperatures, where the Mn can be grown at cold temperatures and the Si cap at higher temperatures.
In this way we attempt to avoid the large Mn precipitates that form during the annealing of submonolayers of Mn deposited on Si(001).\cite{Krause:2006jvsta}
By adding two atomic layers of Pb that surface segregate during growth, the surface diffusion length of Si increases,\cite{Voigtlander:1995prb} which enables the crystalline growth of the capping layer at a lower substrate temperature than is otherwise possible on a clean Si(001) surface.\cite{Evans:1996prb} Lead also has the advantage of being a group IV element that does not act as a dopant in Si and has been successfully used to incorporate As in Si at low temperatures.\cite{Dubon:2001apl}
The presence of the surfactant also modifies the formation energies of the dopants, as found in GaP,\cite{{zhuprl}, {zhujcg}, {lixin}} which could give rise to other important opportunities.

Our previous x-ray absorption fine structure (XAFS) results, however, show that the Mn $\delta$-doped layers also have the strong tendency to form precipitates.\cite{Xiao:2009jpcs} Previously, \cite{Kahwaji:2012prb} we showed that the structure of ultrathin films of Mn on Si is highly sensitive to Mn concentration and to substrate temperature during both the Mn growth and the growth of the Si capping layer. Our XAFS and magnetometry experiments showed that the growth temperature of the Si capping layer influences the local coordination of Mn and leads to MnSi precipitates with B2- and B20-like crystal structures with completely different magnetic properties, and a $T_C$ varying between 10 K to above 250 K.  The B2-MnSi phase was predicted by DFT calculations to have a high $T_C$.\cite{Hortamani:2006prb}

\begin{figure}
\centering
\includegraphics [width=7cm]{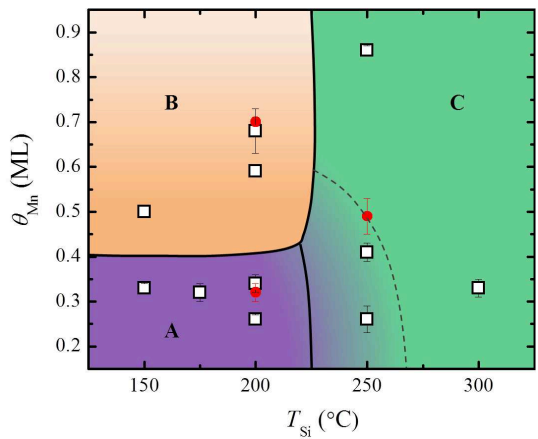}
 \caption{(color online) Growth phase diagram based on Mn coverage $\theta_\textrm{Mn}$ and Si cap growth temperature $\emph{T}_{Si}$. The three control samples without Pb are shown in red circles.  We identify three regions: A, B, and C in the growth diagram based on the magnetic and structural properties of the samples discussed in the text. The dashed line bounds a mixed-phase in region-C. We define 1 ML to correspond to the density of atoms in the Si(001) plane ($6.78 \times 10^{19}$ atoms/m$^{2}$). }
\label{fig:diagram}
\end{figure}

With the complementary data obtained from XAFS, Rutherford backscattering spectroscopy (RBS) and transmission electron microscopy (TEM) measurements, we investigated the structural evolution of the $\delta$-doped layers in a range of Mn coverages $0.26$~ML $\leq \theta_\textrm{Mn} \leq  0.86$~ML and Si-cap growth temperatures $150^{\circ}$C$ ~\leq T_{\textrm{Si}} \leq 300 ^{\circ}$C. The three regions of the growth phase diagram are presented in Fig.~\ref{fig:diagram}, labeled A - C, and the sample descriptions are listed in Table~\ref{tab:samples}.  Superconducting quantum interference device (SQUID) magnetometry measurements show three distinct regions of magnetic behaviors. In region-A, we create a material with $\emph{T}_{C}>$ 400 K  that results from two cooperative phases, where one is a nanocrystalline B2-like MnSi precipitate and the other is possibly from Mn occupying Si substitutional sites (Mn$_{\textrm{Si}}$).
In region-B, ferromagnetism is absent. In region-C, large nano-disks with a B20-like MnSi structure precipitate to form a magnetic phase with an ordering temperature $T_C \approx 170$ K. Density functional theory (DFT) calculations are presented to give insight into the role of the surfactant. Lead is found to decrease the formation energy of Si vacancies, and thereby create a mechanism for increasing the concentration of substitutional Mn.

\section{Methods}

\subsection{Growth}

We used MBE to grow Mn $\delta$-doped films on low (1-20 $\Omega$ cm) and high resistivity (600-1200 $\Omega$ cm) boron-doped Si(001) substrates. The Si wafers were prepared following the method of Ref. \onlinecite{Kahwaji:2012prb}.
For all samples, a 20 nm Si buffer layer was deposited onto the ($2\times1$) surface-reconstructed Si(001) prior to thin film growth. We deposited 2 ML Pb, measured by reflection high-energy electron diffraction (RHEED) intensity oscillations, at room temperature. By the completion of the second layer, a ($2\times1$) reconstruction of the Pb covered surface was observed again.\cite{Zhao1992, Li:1994PRB}

\begin{table}[h]
\caption{ Samples and their corresponding $\theta_\textrm{Mn}$, $T_{\textrm{Si}}$ and region in growth-diagram. We refer to the samples in the text as listed in the first column.}
\label{tab:samples}
\begin{ruledtabular}
\begin{tabular}{lccc}
\textbf{Sample} &\textbf{ $\theta_\textrm{Mn}~(ML)$} & \textbf{$T_{\textrm{Si}}~(\celsius)$} & \textbf{region}\\ \hline
\textbf{$a_1$} & 0.26 & 200 & A\\ 
\textbf{$a_2$} & 0.32 & 200 & A \textbf{(No Pb)}\\ 
\textbf{$b_1$} & 0.59 & 200 & B\\ 
\textbf{$b_2$} & 0.68 & 200 & B\\
\textbf{$b_3$} & 0.7 & 200 & B \textbf{(No Pb)}\\
\textbf{$c_1$} & 0.26 & 250 & C\\
\textbf{$c_2$} & 0.41 & 250 & C\\
\textbf{$c_3$} & 0.49 & 250 & C \textbf{(No Pb)}\\
\textbf{$c_4$} & 0.86 & 250 & C\\
\textbf{$c_5$} & 0.33 & 300 & C\\
\end{tabular}
\end{ruledtabular}
\end{table}

The substrate temperature was lowered to $T <0~\celsius$ before the deposition of a submonolayer of Mn. The Mn-flux rate was monitored during growth with a Bayard-Alpert ionization gauge placed in the Mn beam, and the amount of deposited Mn was subsequently confirmed with RBS. \emph{Ex-situ} x-ray photoelectron spectroscopy measurements confirm that Pb segregates to the surface during the growth of the 12.4 nm-thick Si capping layer that was deposited onto a substrate held at a temperature $T_{\textrm{Si}}$.    The RHEED diffraction pattern showed that the cap remained crystalline throughout the growth, although TEM shows stacking faults in the capping layer in samples in region-B and region-C. We also prepared control samples without a Pb-layer in order to determine how the Pb affects the Mn incorporation and the magnetic properties.

\subsection{ RBS }

\begin{figure}[h]
  \centering
  \includegraphics[width=7.5cm]{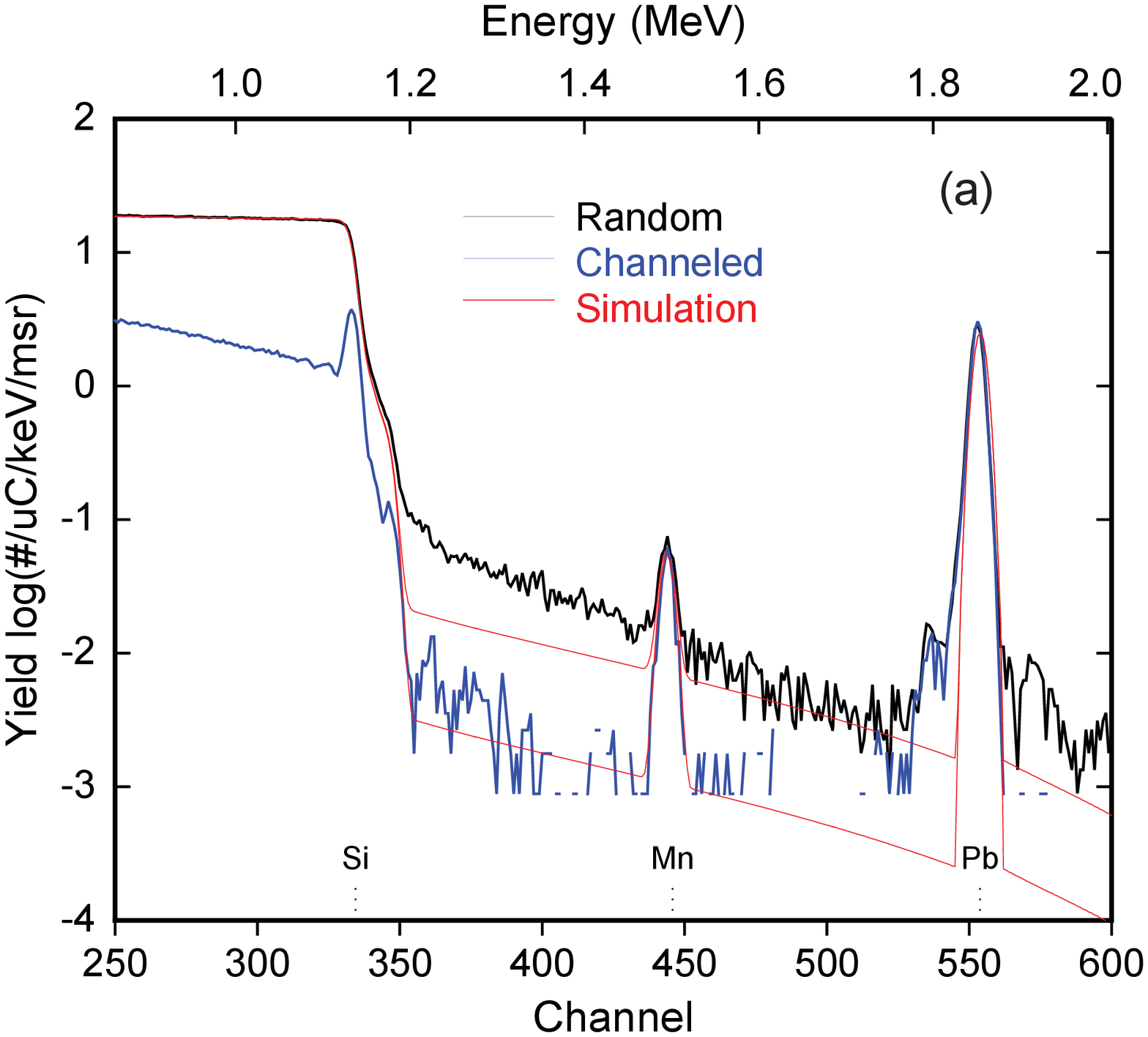}
  \includegraphics[width=7cm]{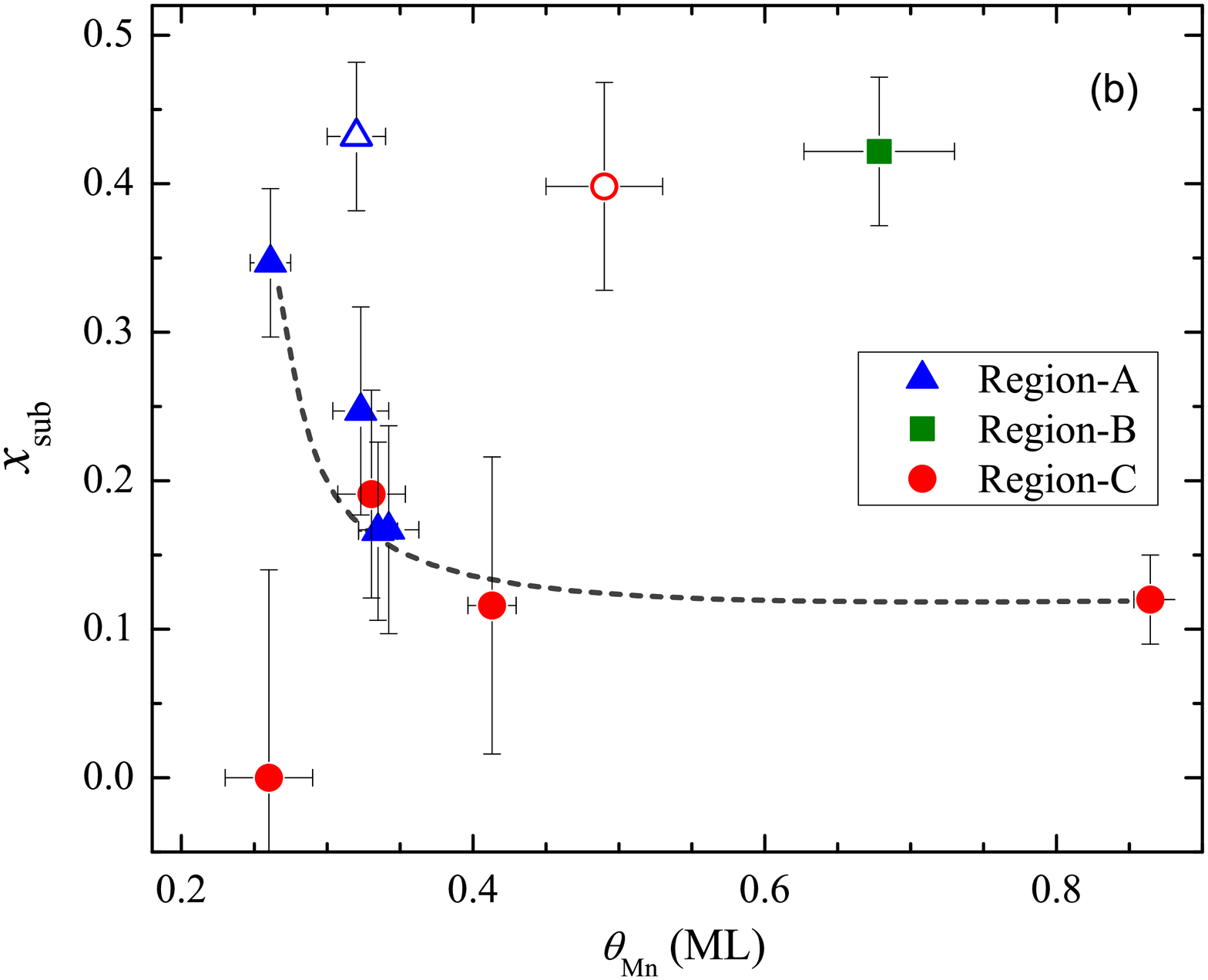}
  \caption{(color online) (a) RBS spectra of one of the samples collected in random and channeled orientations. The Mn substitutional fraction $x_{\textrm{sub}}$ is calculated by using the difference between the areas of Mn peaks in both orientations. (b) Variation of $x_{\textrm{sub}}$ with Mn coverage $\theta_\textrm{Mn}$ for samples in all three regions of the growth diagram. Open triangle (circle) refers to control sample of region-A (region-C) grown without Pb and the dashed line is a guide to the eye.}
  \label{fig:RBSRatios}
  \end{figure}

We determined the Mn-dopant concentration and the fraction of Mn atoms that occupy Si substitutional sites ($x_{\textrm{sub}}$) from RBS.  The 2 MeV He ions were directed along the [100] direction of the sample.  A wide angle detector (WAD) measured the ion yield backscattered through an angle of $\approx 170\degree$, and a glancing angle detector (GAD) collected ions with grazing incidence. Since the WAD measurements are less prone to small changes in geometry, we used those values to calculate the amount of Mn and $x_{\textrm{sub}}$, whereas the values obtained from GAD were employed to determine the depth of the Mn layer. We estimated  $x_{\textrm{sub}}$ from the difference in the Mn signal measured in random and channeling orientations. Usually, the Mn concentrations determined from random orientation spectra are more reliable than the Mn amounts determined from channeling orientation because they are compared to the Si random height, whereas the channeled spectra depend on less reliable charge normalization to extract the Mn concentration. Therefore we checked the values of $x_{\textrm{sub}}$ by using the signal from Pb as a reference, assuming that the Pb peak does not allow channeling. As a further check on the reliability of charge integration, and specifically for control samples without Pb, we verified that the minimum ion yield measured from the spectrum height from ions scattered deep in the Si was within the expected range of 4-5\% for a well channeled spectrum. Furthermore, since in a channeling experiment the incident ions tend to travel in the center of the channel, the probability of backscattering from interstitial Mn is enhanced, this is referred to as \emph{flux peaking}. Because of flux peaking, the channeled signal overestimates the amount of interstitial Mn and, consequently, the calculated substitutional Mn fractions should be interpreted as lower thresholds values.\cite{Andersen:1971RE}
Characteristic RBS spectra measured in random and channeled orientations and the variation of $x_{\textrm{sub}}$ with $\theta_\textrm{Mn}$ are shown in Fig.~\ref{fig:RBSRatios}.

\subsection{XAFS and TEM}
We probed the local coordination of Mn by XAFS in fluorescence mode. Polarization-dependent XAFS spectra were collected at the Mn K-edge with the samples mounted on a spinner turning near 30 - 40 Hz to minimise contributions from Bragg peaks. X-rays were incident on the sample at an angle between 5 and 7$\degree$~above the plane of the wafer (83 - 85$\degree$ off the surface/substrate normal). Hence, for out-of-plane (oop) polarization measurements, the electric field vector of the linearly polarized X-rays was then oriented between 5 - 7$\degree$ off the [001] substrate normal direction of the spinning wafer. In-plane (ip) polarization measurements were averaged over all in-plane orientations perpendicular to [001]. Given the high symmetry of the [001] direction in Si, no orientation-dependence was expected in the in-plane XAFS signal.
The XAFS experiments and analysis were conducted as outlined in Ref. \onlinecite{Kahwaji:2012prb} using the facilities of the PNC/XSD ID beamline \cite{Heald1999JSR} at the Advanced Photon Source, sector 20.

\begin{figure}[h]
  \centering
  \includegraphics [width=\columnwidth]{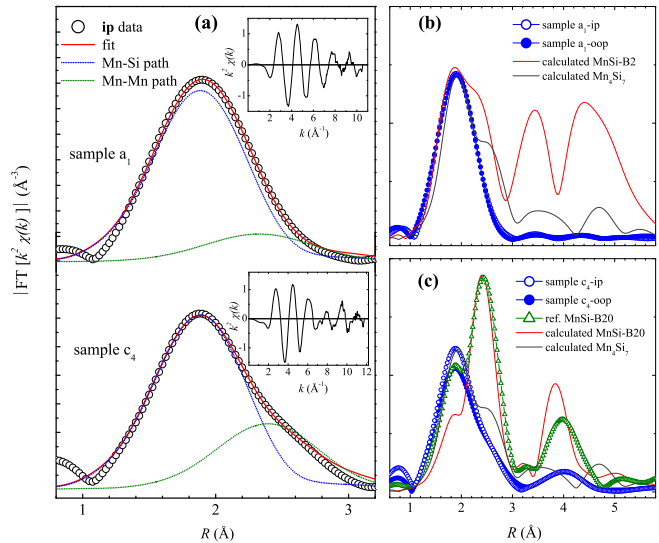}
  \caption{(color online) (a) In-plane XAFS data (insets) and corresponding fit in \emph{R}-space for samples $a_1$ and $c_4$. Profile functions of Mn-Si and Mn-Mn scattering paths are shown for each sample and the fitting range is within $R = 1.40-2.82$ \AA. In (b) and (c) we compare the $|FT [k^2~\chi (k)]|$ of samples $a_1$ and $c_4$ with the calculated $|FT [k^2~\chi (k)]|$ of MnSi-B2, MnSi-B20 and Mn$_4$Si$_7$ structures. We used the same window function and similar $k$-range for the FT of the calculated structures and scaled their $|FT [k^2~\chi (k)]|$ to match either the Si (MnSi-B2 and Mn$_4$Si$_7$) or the Mn (MnSi-B20) peak of the measured spectra.}\label{fig:XAFS}
\end{figure}

We obtained the coordination parameters from fitting the magnitude of the Fourier transform (FT) of the $k^2$-weighted XAFS interference function $\chi(k)$, $|FT [k^2~\chi (k)]|$, in \emph{R}-space. We show two XAFS $\chi(k)$ in Fig.~\ref{fig:XAFS} that are representative of our XAFS data. As can be seen from the insets of Fig.~\ref{fig:XAFS}(a), there is an increase in the noise level of $k^2~\chi (k)$ for $k > 10.0$~\AA$^{-1}$ therefore, we limited the FT to within the closest zero-crossing of the \emph{k}-range $= 2.3 - 10.0$ \AA$^{-1}$. We included two scattering paths that corresponded to the first Mn-Si shell and the second Mn-Mn shell, with the exception of sample $b_1$-ip, which had two separate Mn-Si shells. To further confirm the composition of the two shells, we applied the equations derived from \emph{beats theory} in Refs. \onlinecite{{Rudolf},{Bunker}} to calculate the difference in radii between the first two shells, $\Delta R = R_2 - R_1$. The values we found were consistent with the $\Delta R$ that we obtained from our best fit values of $R_1$ and $R_2$.

\begin{table*}
\caption{Coordination numbers ($N$) and distances ($R$) obtained from the fit of the in-plane (ip) and out-of-plane (oop) XAFS data of the samples and the ip data of a 5 nm MnSi-B20 film. Estimated parameters of a tetragonally distorted \emph{t}-MnSi-B2 structure and of Mn in Si substitutional site, $Mn_{Si}$, are also included. $A_{Mn-Mn}$ is the area under the Mn-Mn profile function relative to that of Mn-Si. The percent residual ($R\%$) and uncertainty of each fitting parameter (in parentheses) are also included. The energy shift $\Delta E_0 = 5.9 \pm 0.6$ eV and $S_{0}^{2} = 0.70 \pm 0.02$ were obtained from the reference MnSi-B20 sample and treated as fixed parameters in the $\delta$-doped samples.}
\label{tab:XAFS}
\begin{ruledtabular}
\begin{tabular}{lcccccccr}
 \textbf{Samples}&$N_1~(Si)$&$R_1$~(\AA)&$\sigma _1^{2}$~(\AA$^2$)&$N_2~(Mn)$&$R_2$~(\AA)&$\sigma _{2}^{2}$~(\AA$^2$)&$A_{Mn-Mn}~(\%)$&$R~(\%)$\\ \hline
\textbf{$a_1$-ip}&$6.7(7)$&$2.37(1)$ &$0.009(2)$&-&$2.76(2)$&-&$18.8$&$1.04$\\
\textbf{$a_1$-oop}&$7(1)$&$2.378(5)$ &$0.010(2)$&-&$2.75(3)$&-&$15.0$&$1.14$\\ \hline
\textbf{$b_1$-ip}&$2(1)$&$2.22(3)$ &$0.006$&$7(1)$\textbf{(Si)}&$2.42(2)\textbf{(Si)}$&$\sigma _{1}^{2}$&-&$2.61$\\
\textbf{$b_1$-oop}&$6.1(8)$&$2.38(1)$ &$0.008(2)$&-&$2.80(1)$&-&$21.4$&$0.91$\\ \hline
\textbf{$b_3$-ip (No Pb)}&$7.7(4)$&$2.365(5)$ &$0.0096(8)$&-&$2.79(1)$&-&$9.2$&$0.47$\\
\textbf{$b_3$-oop (No Pb)}&$7.0(6)$&$2.367(4)$ &$0.008(1)$&-&$2.82(2)$&-&$13.2$&$0.82$\\ \hline
\textbf{$c_4$-ip}&$6.3(6)$&$2.38(1)$ &$0.010(2)$&-&$2.817(5)$&-&$41.0$&$1.11$\\
\textbf{$c_4$-oop}&$6.3(6)$&$2.38(1)$ &$0.012(1)$&-&$2.829(4)$&-&$44.9$&$0.62$\\ \hline \hline
\textbf{MnSi-B20}&$6.5$&$2.39(1)$ &$0.010(4)$&$5.9$&$2.788(6)$&$0.005$&-&$1.88$\\ \hline
\textbf{\emph{t}-MnSi-B2}&$8$&$2.370$ & &$6$&$2.790$& & & \\ \hline
\textbf{Mn$_{Si}$}\footnote{We consider 1 substitutional Mn atom in the Si lattice since $\theta_\textrm{Mn} < 1$~ML. In this case, the second shell corresponds to Mn-Si with $R_2 = 3.84$~\AA, which is outside our fitting range and therefore not included in the table.}&$4$&$2.351$ & & & & & & \\
\end{tabular}
\end{ruledtabular}
\end{table*}

In Table~\ref{tab:XAFS}, we report the XAFS parameters of the $\delta$-doped samples. For comparison, we also include the parameters of a 5 nm MnSi/Si(111) reference sample with a B20 crystal structure, as well as the parameters expected for Mn in a Si substitutional site and for a Mn in a tetragonally distorted B2 (CsCl) structure. The amplitude reduction factor $S_{0}^{2} = 0.70$ and the energy shift $\Delta E_{0} = 5.9$ eV were fixed to the values obtained from the reference MnSi-B20 sample. Given the weak Mn second shell and the strong correlation between $N_2$ and the mean square relative displacement $\sigma _{2}^{2}$ for the measured samples, we could not determine a value of $N_2$. Instead, we refer to the area under the Mn-Mn profile function, $A_{Mn-Mn}$, to represent the relative combined change in $N_{2}$ and $\sigma _{2}^{2}$.

 We further characterized the structure of the samples with plan view and cross-sectional TEM, imaged with 300 kV Philips CM30 TEM.  Samples were prepared by low-angle mechanical polishing and subsequently cleaned with Ar-ion milling.\cite{Robertson:2006ws}
We also used energy dispersive x-ray spectroscopy (EDX) to probe the chemical composition of the precipitates that formed in certain samples.

\subsection{Magnetometry}
We analyzed the magnetic properties of the samples by measuring their moment as a function of field (\emph{m-H}) and remanent moment as a function of temperature ($m_r(T)$) with a Quantum Design MPMS XL-5 SQUID magnetometer. For all magnetic measurements, 3-5 pieces of each sample were mounted inside a sample holder consisting of a clear plastic straw. We did not find a difference in the magnetization when measured with the field applied parallel or perpendicular to the [110] orientation of the films, which is likely due to disorder in the Mn layer. To remove the paramagnetic and diamagnetic responses of the  low-resistivity Si(001), where the boron concentration is of  the order of $10^{15}-10^{16}$ atoms$/$cm$^{3}$, we measured the magnetic response of a control sample that consisted of a 100 nm Si buffer layer grown on Si(001) and subtracted it from that of the $\delta$-doped samples. We could not detect a paramagnetic signal from a similar control sample grown on high-resistivity Si(001), and therefore only the diamagnetic response of the Si was subtracted from the samples grown on the high-resistivity substrates.  Furthermore, we measured the magnetic response of a sample prepared with 2 ML Pb under the same conditions of  the $\delta$-doped films but without Mn. We found that the saturation and remanent moments of this Mn-free control sample are zero within the error of the measurement, which confirms that the remanent magnetization in the $\delta$-doped films arises from Mn.

We calculated the Mn saturation moment from the \emph{m-H} loops measured at $T = 2$~ K. For $m_r(T)$ measurements, we saturated the magnetization in a 5 T applied field, and then to ensure that the field is set to zero, we used the magnet reset option of the SQUID magnetometer to discharge the magnetic flux trapped in the coils. Measurements of $m_r(T)$ were then made while increasing the temperature from 2~K.  We estimated the Curie temperature $T_C$ from the point where $m_r(T)$ drops to zero.

\section{Magnetic and Structural Properties}

\begin{figure}[h]
 \centering
 \includegraphics[width=\columnwidth]{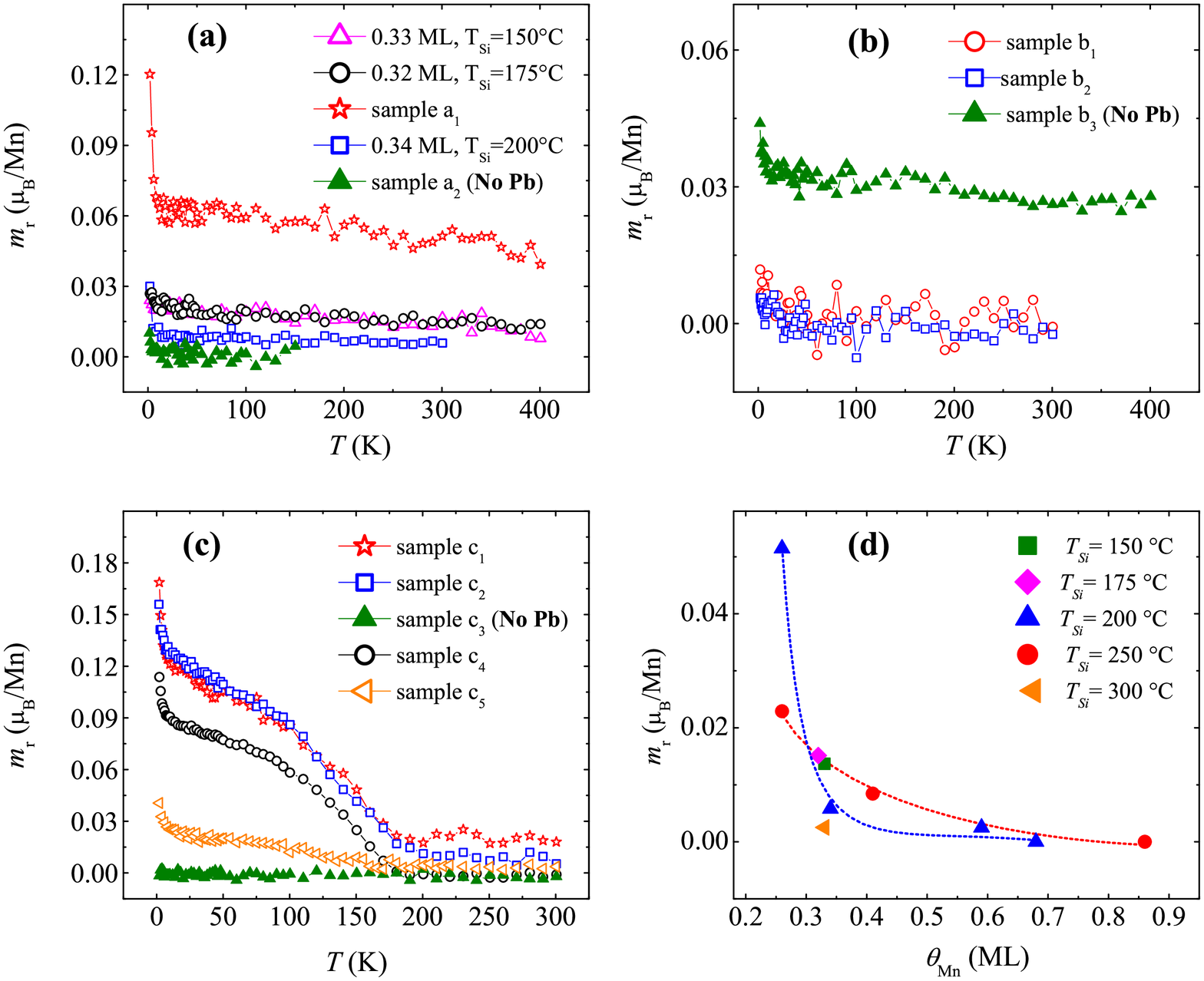}
  \caption{(color online) (a) The non-zero $m_{r}(T)$ indicate a ferromagnetic phase with $T_C > 400$~K for samples in region A. The control sample of this region (filled triangles) has zero remanent moment. (b) $m_r(T)$ show no ferromagnetic ordering for samples grown with Pb in region B but a $T_C > 400$~K for the control sample. (c) A magnetic phase transition is observed in region-C with $T_C \simeq 170$~K for all samples with Pb. There is no remanent moment for the control sample of this region. (d) Remanent moment measured at $T = 250$~K as a function of $\theta_\textrm{Mn}$ for all $T_{\textrm{Si}}$. Dashed lines connect the samples grown with same $T_{\textrm{Si}}$.}
\label{fig:mrALL}
\end{figure}

\begin{figure}[!]
  \centering
  \includegraphics[width=4cm]{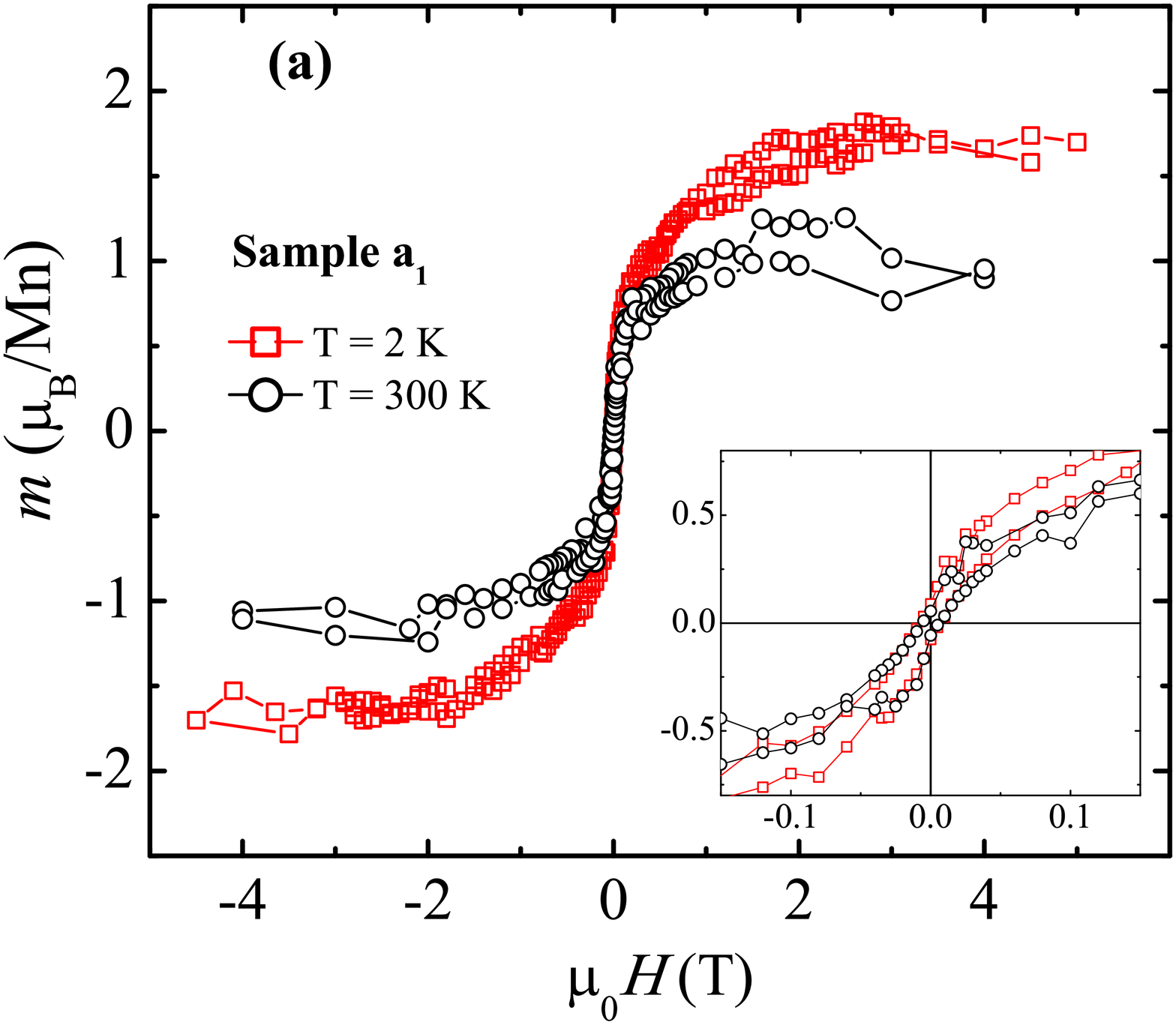}
  \quad
  \includegraphics[width=4cm]{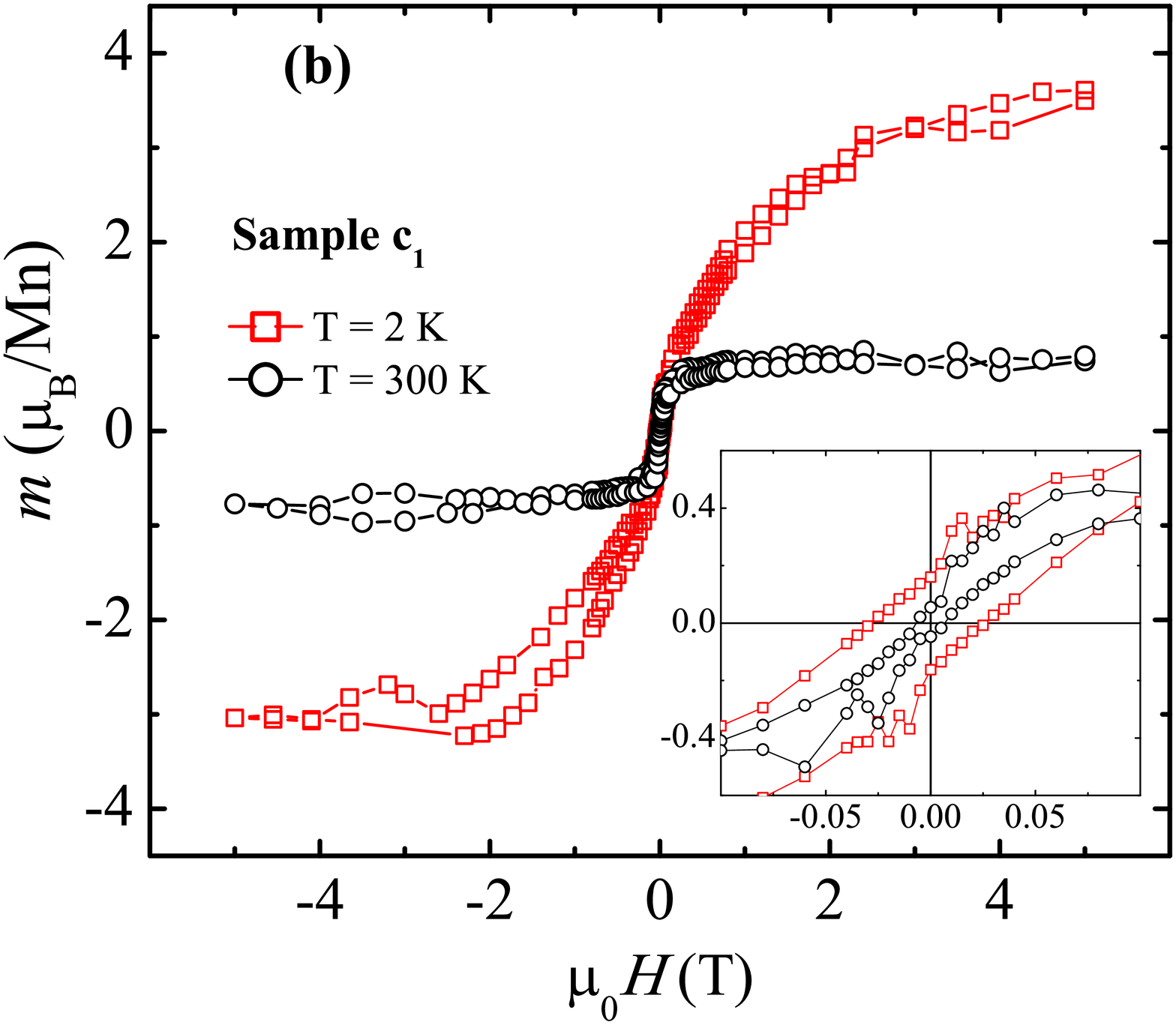}
  \centering
  \includegraphics[width=4cm]{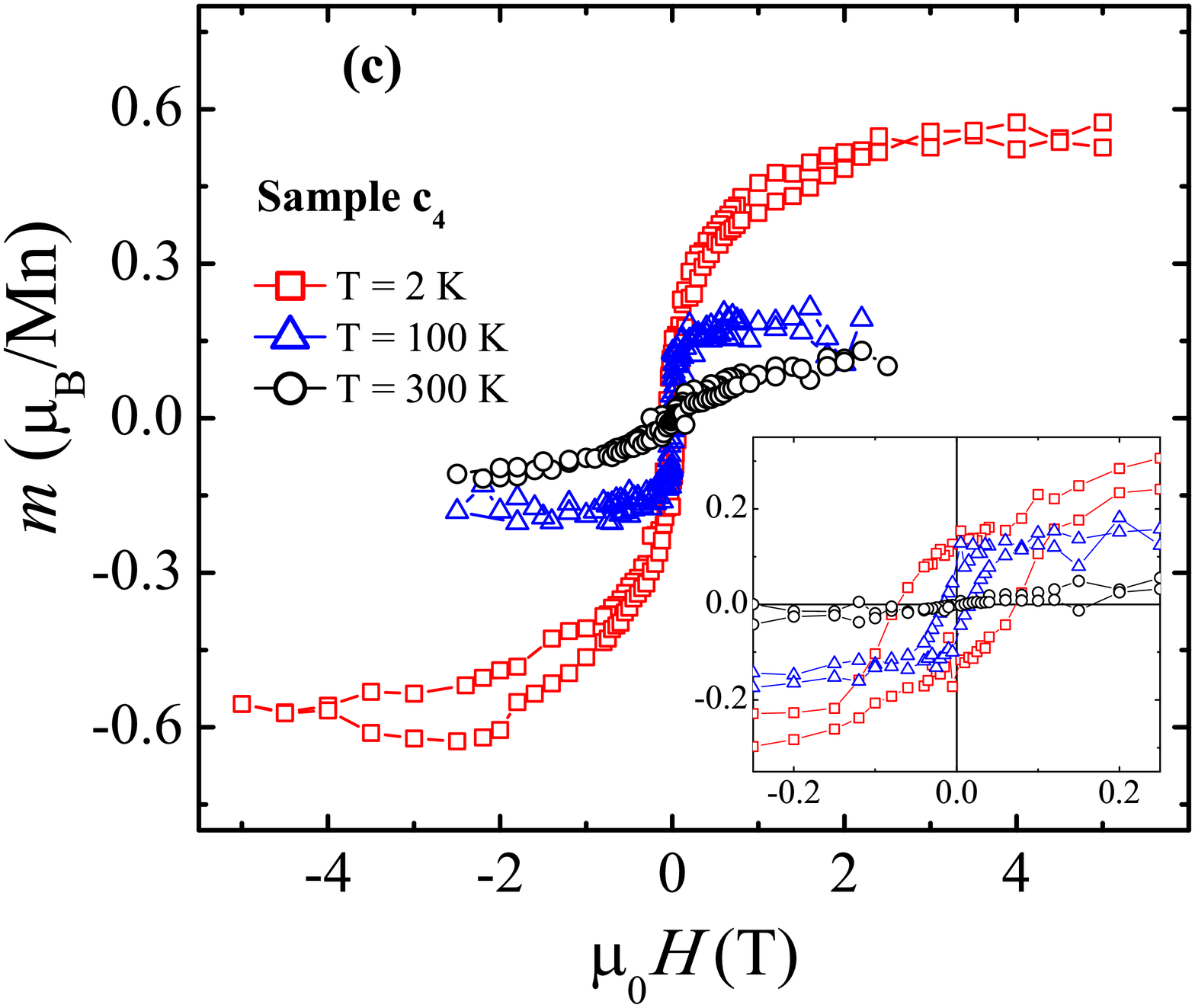}
  \quad
  \includegraphics[width=4cm]{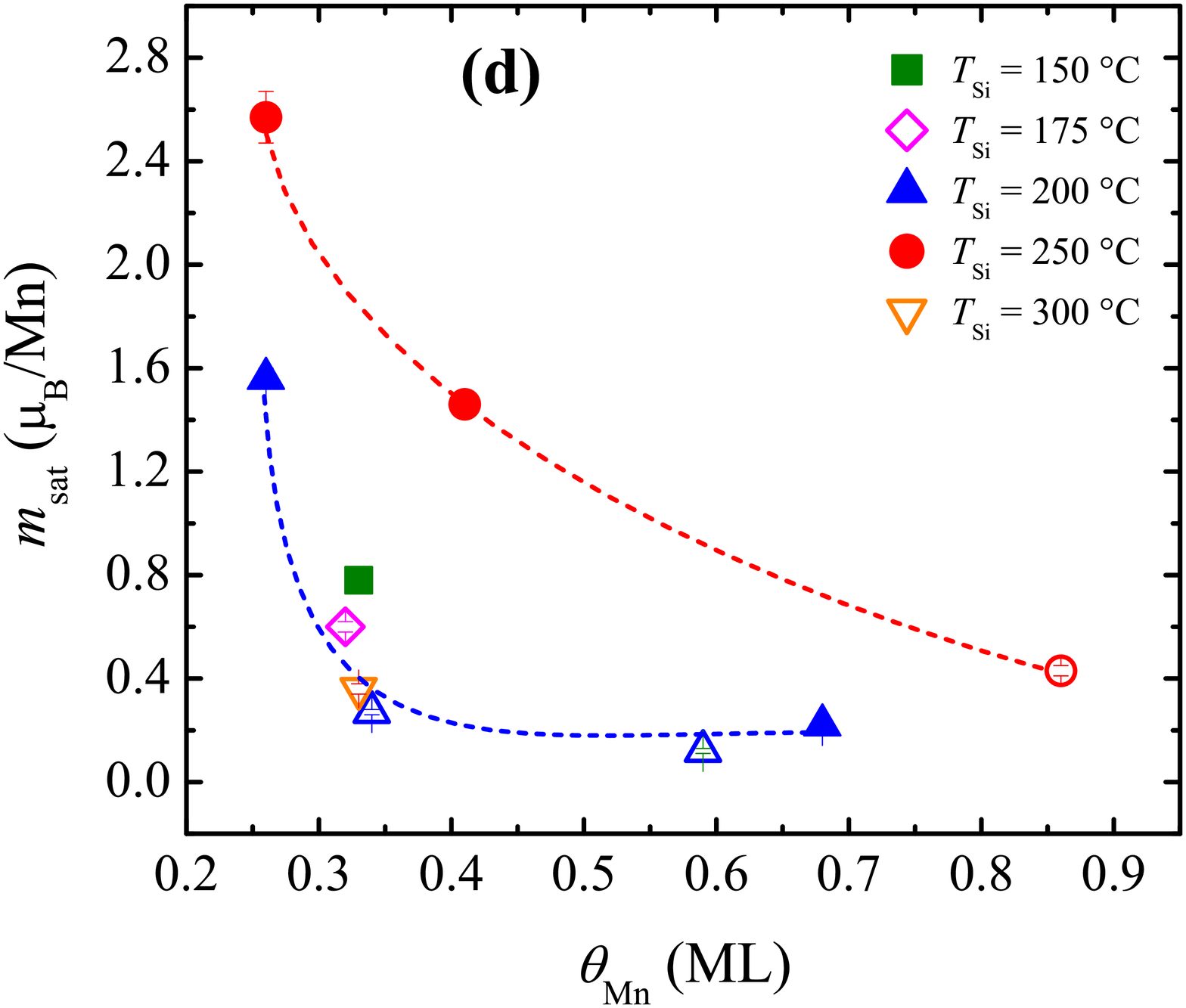}
  \caption{ (color online) $\emph{m-H}$ loops measured at low and high temperatures as indicated. The insets show that there is a hysteresis at $T = 300$~K for samples (a) $a_1$ and (b) $c_1$ but it vanishes for sample $c_4$ (c). (d) Variation of Mn saturation moment $m_{sat}$ calculated at $T = 2$~K as a function of $\theta_\textrm{Mn}$. Open symbols refer to samples grown on low resistivity Si(001) where the B paramagnetic contribution has been subtracted.}
\label{fig:mH-msat}
\end{figure}

SQUID results demonstrate distinct magnetic behaviors in each of the three regions of the growth-phase diagram shown in Fig.~\ref{fig:diagram}.  Region-A has a $T_C$ well above room temperature, as indicated by the non-zero remanent moment measured up to $T = 400$~K (Fig.~\ref{fig:mrALL}(a)). The existence of magnetic order is confirmed by the presence of hysteresis shown in Fig.~\ref{fig:mH-msat}(a). Sample $a_1$ was found to have the optimal growth conditions that produced the largest $m_r$ (Fig.~\ref{fig:mrALL} (a)) and the largest saturation moment, $m_{sat} = 1.56~\mu_B$/Mn ($\mu_B$ thereafter) measured at $T=2$~K (Fig.~\ref{fig:mH-msat}(d)). This sample has the lowest Mn concentration, $\theta_\textrm{Mn} = 0.26$~ML, and a growth temperature that is closest to the phase boundary between regions-A and C. The $m_r$ decreases as $T_{\textrm{Si}}$ drops from this optimal $T_{\textrm{Si}} = 200~\celsius$  (Fig.~\ref{fig:mrALL}(d)), while the $m_{sat}$ measured at $T = 2$~K drops quickly with increasing $\theta_\textrm{Mn}$ (Fig.~\ref{fig:mH-msat}(d)).

Fits to the XAFS spectra resolve the first two shells surrounding the Mn atoms that correspond to the Si nearest neighbors and Mn next-nearest neighbors (see Table~\ref{tab:XAFS}).  The Mn-Si bond length, $R_1 = 2.374$~\AA~ and the coordination number $N_1 = 6.7 \pm 0.7$ are larger than expected for substitutional Mn (see Table~\ref{tab:XAFS}). The large coordination indicates the formation of precipitates, although these phases could not be detected in our TEM experiments. A qualitative comparison between XAFS measurements of sample $a_1$ and calculated spectra shows that the local Mn environment does not correspond to either MnSi-B20, MnSi$_{1.7}$ or Mn$_5$Si$_3$ crystal structures that are commonly produced from deposition of Mn on Si(001).\cite{Lian:1986apl, Lippitz}  Furthermore, the Mn-Mn bond length $R_2 =2.75$~\AA~ is significantly contracted relative to that measured for MnSi$_{1.7}$ precipitates in Ref.~\onlinecite{Orlov:2009jetp}.

According to DFT predictions, tetragonally distorted MnSi-B2 is the most energetically favorable structure that forms on Si(001), since MnSi-B2 phase has a reasonably good lattice match to Si(001).\cite{HortamaniPRB78}  We reported the existence of a B2-like phase in our previous work of MnSi films on Si(001).\cite{Kahwaji:2012prb} A comparison between the fitting parameters in Table~\ref{tab:XAFS} with those expected for a B2 show that this structure is the closest match to sample $a_1$. The expected 8-fold coordinated crystal structure can be constructed from a Si lattice by placing Si and Mn into interstitial sites. Missing interstitials would account for some of the discrepancy between the measured and the expected coordination number for a B2 structure.  RBS, however, shows that only 65\% of the Mn in sample $a_1$ is in interstitial sites, which indicates that Mn in this sample has at least two distinct environments.  Although the XAFS is not inconsistent with the co-existence of both a B2 phase and Mn in Si substitutional sites (Mn$_{\textrm{Si}}$) such a fit would introduce too many fitting parameters to permit a quantitative analysis.

The SQUID data is also suggestive of two components: Fig.\ref{fig:mH-msat}(a) reveals a wasp-waisted hysteresis loop that can be constructed from the superposition of a ferromagnetic and a superparamagnetic hysteresis loop, as shown in the case of MnSi$_{1.7}$ nanoparticles in Si.\cite{Zhou:2009prb} The ferromagnetic hysteresis loop seen at low magnetic field in the inset of Fig.~\ref{fig:mH-msat}(a) has a weaker temperature dependence than the s-shaped portion of the $m-H$ curve visible at high fields, which suggests that these components have different magnetic behaviors.

\begin{figure}
  \centering
  \includegraphics[width=\columnwidth]{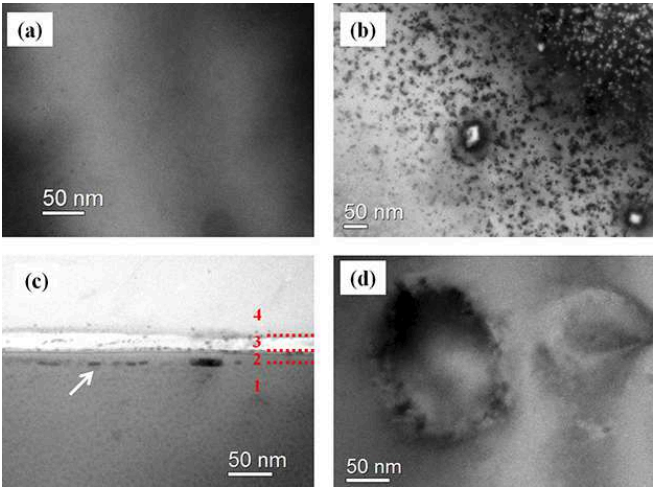}
  \caption{
   (a) The bright field plan-view TEM image of sample $a_1$ hardly shows any precipitates whereas (b) Mn-rich precipitates are clearly observed in sample $b_2$. (c) Cross-section TEM image of sample $c_4$ shows Mn-rich precipitates (dark features indicated by arrow) distributed at the interface between the substrate (1) and the Si-cap (2).  The epoxy layer (3) that glues the samples to a second Si wafer (4) is also visible in the image. The precipitates in this sample have nano-disk shape as seen in (d)}\label{fig:TEM}
\end{figure}

The samples in Region B of the growth-phase-diagram have $T_C$'s near zero (Fig.~\ref{fig:mrALL}(b)) and the $m_{sat}$ values are small (Fig.~\ref{fig:mH-msat}(d)).
However, there is a noticeable change in the Mn environment when $\theta_\textrm{Mn}$ increases to $0.59$~ML at $T_{\textrm{Si}} = 200~\celsius$. In particular, we note the distinctive in-plane structure of sample $b_1$ which has two Mn-Si shells as opposed to Mn-Si and Mn-Mn shells for all other samples. The short Mn-Si distance $R_1 = 2.22$~\AA~ forms only in plane.
The TEM images of Fig.~\ref{fig:TEM}(a) show that sample $b_2$ contains precipitates with an estimated lateral size of up to 15 nm. Curiously, samples from this region still show a considerably large $x_{\textrm{sub}}$ (Fig.~\ref{fig:RBSRatios}).  The small $m_{sat}$ and low $T_C$ indicate that these sites are not magnetically active.

In region-C of Fig.~\ref{fig:diagram}, the $m_r$ has a concave shape typical of ferromagnetism that drops either to zero or to a small constant value at a temperature of  170~K. This transition temperature is unaffected by a variation in $\theta_\textrm{Mn}$ or an increase in  $T_{\textrm{Si}}$  to $300~\celsius$ (Fig.~\ref{fig:mrALL}(c)), which reflects a higher stability relative to the phase in region-A.  It is likely that this magnetic behavior arises from the large precipitates that form in this region and can be seen in the cross-section TEM image of sample $c_4$ (Fig.~\ref{fig:TEM}(c)), given that the onset of this new low temperature feature coincides with their appearance. The nano-disk shaped precipitates (Fig.~\ref{fig:TEM}(d)) distribute themselves at the interface with a range of sizes that vary between 10-200 nm in diameter.\\
Energy-dispersive x-ray spectroscopy (EDX) showed that the precipitates contain Mn, but it was not possible to determine their phase from the weak reflections in the selected area diffraction pattern (SADP) measurements. A comparison between the XAFS function of sample $c_4$ and that for the MnSi-B20 structure shows similar features up to $R = 5$~\AA (Fig.~\ref{fig:XAFS}(c)). Although the first Mn-Si shell coordination number and radius are within the error limits of those of MnSi-B20, the ip and oop bond lengths of the second shell appear to be expanded relative to the $R_2$ of a B20 structure (see Table~\ref{tab:XAFS}).  Furthermore, the transition at $T = 170$~K is not consistent with the $T_C \le 44$~K of ordered MnSi-B20 thin films.\cite{KarhuPRB82} We expect that the  low $\theta_{\textrm{Mn}}$ and low $T_{\textrm{Si}}$ create a defected MnSi-B20 structure. Such structures have been observed in similar Mn-doped Si systems.\cite{{WolskaPRB75},{Kahwaji:2012prb}}

The presence of a second phase is evidenced by the non-zero $m_r$ above 170~K  in the lower $\theta_\textrm{Mn}$ samples in region-C.  The $m_r$ of sample $c_1$ above 170~K matches the temperature dependence of the remanent moment of region-A samples.  Furthermore, the values of the high-temperature $m_r$ for the region-C samples are comparable to those of region-A (see Fig.~\ref{fig:mrALL}(d)) with the exception of the optimal sample $a_1$. This suggests that while the higher $T_{\textrm{Si}}$ drives the Mn into larger precipitates with $T_C \approx 170$~K, some of the Mn forms the same structure found in region-A.  Like the phase in region-A, this second phase is sensitive to $\theta_\textrm{Mn}$.
Figure~\ref{fig:mrALL}(c) shows that this second phase is absent from sample $c_4$ ($\theta_\textrm{Mn} = 0.86$~ML).  It is also sensitive to growth temperature: $m_r(T) = 0$ above $T = 170$~K for the $T_{\textrm{Si}} = 300~\celsius$ sample in Fig.~\ref{fig:mrALL}(c).

\begin{figure}
  \centering
  \includegraphics[width=7cm]{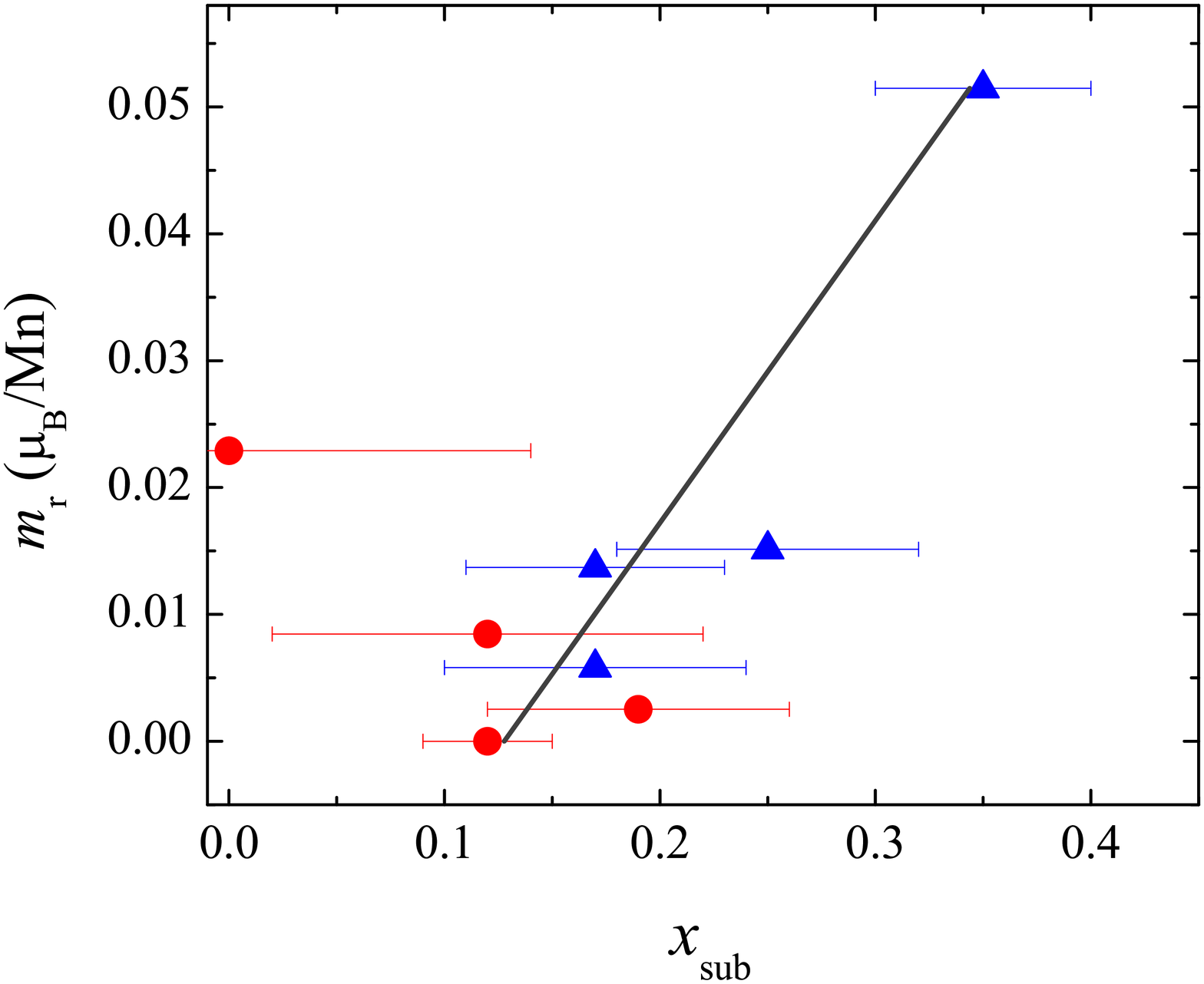}
  \caption{(color online) Variation of remanent moment measured at $T = 250$ K with the fraction of substitutional Mn, $x_{\textrm{sub}}$. A possible correlation exists for samples of region-A (blue triangles) and region-C (red circles), as shown by the black line.}\label{fig:MRvsXsub}
\end{figure}

The contrasting magnetic behavior of samples in the three regions of the growth-phase diagram raises the question of the mechanism of the magnetic order.
For regions-A and C, the $m_r$ and the $m_{sat}$ are found to decrease with increasing $\theta_\textrm{Mn}$, as shown in Figs.~\ref{fig:mrALL} (d)  and~\ref{fig:mH-msat}(d).  As the concentration of Mn increases, so does the probability of the formation of precipitates. It is interesting that $x_{\textrm{sub}}$ also follows this trend (Fig.~\ref{fig:RBSRatios}) for regions-A and C.
Given the nanoscale size of the precipitates, however, it is important to understand the role that the Si matrix must play in mediating the ferromagnetic order.
One possible mechanism is the formation of a DMS in the regions between the precipitates. RBS measurements show that the remanent magnetization is possibly correlated with the fraction of substitutional Mn, $x_{\textrm{sub}}$. In Fig.~\ref{fig:MRvsXsub}, the $m_r$ increases approximately linearly with $x_{\textrm{sub}}$ for samples in regions-A and C. RBS measured $x_{\textrm{sub}}$ = 0 for sample  $c_1$, although flux peaking may be underestimating this value, and the uncertainty in the measurement places the value 1.5 standard deviations from the line of best fit.  Since MnSi-B2 samples are known to be ferromagnetic with a high $T_C$,\cite{Kahwaji:2012prb,HortamaniPRB78} a Si$_{1-x}$Mn$_{x}$ DMS  provides a plausible explanation for how the precipitates in region-A are correlated in order to yield a high $T_C$. A similar mechanism of DMS mediated interaction explains the coupling between the Mn$_x$Ge$_{1-x}$ nanocolumns in Ref.~\onlinecite{Li:2007PRB}, and was shown to compete with an antiferromagnetic dipolar coupling between the nanocolumns.

In region-C, however, the MnSi-B20 precipitates were not expected to have a $T_C$ as large as 170~K due to the small $T_C \leq 44$~K found in MnSi thin films.\cite{Karhu:2012prb} It is possible that the model put forward in Ref.~\onlinecite{Menshov:2011prb} to explain the high $T_C$ in nanocrystalline MnSi$_{1.7}$ precipitates in Si also explains the behavior of the region-C samples.  Like MnSi$_{1.7}$, MnSi-B20 is a weak itinerant magnet.  The defects suggested by the XAFS data may create large localized magnetic moments that are coupled by spin-fluctuations in the MnSi, as proposed by Men'shov et al.\cite{Menshov:2011prb}  A DMS may be the second phase that is mediating the interaction between these large moment precipitates. The explanation could also be related to the interfacial Mn, which is shown theoretically and experimentally to have an enhanced magnetic moment.~\cite{KarhuPRB82, HortamaniPRB78}. It is important to note that, unlike a similar structure of Mn doped Ge where the Mn acts as a surfactant and floats to subsurface sites in the Ge cap \cite{Zeng:2008PRL}, our measurements show that Mn in the $\delta$-doped Si layers remains localized at the interface. For instance, we could not detect Mn-rich inclusions in the Si cap from the cross-sectional TEM image of Fig.~\ref{fig:TEM}(c), but we only observed the MnSi nano-disks at the interface. Also, the depth of the backscattering Mn atoms measured by RBS GAD detector was consistent with the thickness of the Si capping layer ($\sim$ 12.4 nm), which indicates that Mn is buried at the interface.

While Fig.~\ref{fig:MRvsXsub} shows a possible correlation between $m_{r}$ and $x_{\textrm{sub}}$, we observed no correlation between $m_{sat}$ and $x_{\textrm{sub}}$ that would allow us to extract the magnetic moments of the individual magnetic phases in our composite samples.  A difficulty with the interpretation of the RBS data is that we cannot distinguish the substitutional-like Mn in the precipitates from the Mn$_{\textrm{Si}}$.

Surprisingly, the region-B samples were found to have the highest $x_{\textrm{sub}}$ (Fig.~\ref{fig:RBSRatios}(b)). The absence of ferromagnetism in the region-B samples could be due to a number of possibilities. One possibility is the presence of an antiferromagnetic component.  DFT calculations of Mn interstitial-substitutional complexes, however, predict ferromagnetic behaviour.\cite{Bernardini:2004apl} More recent calculations of $\delta$-doped layers of mixed substitutional and interstitial Mn finds a complex non-collinear magnetic order.\cite{Otrokov:2011prb}  A second possibility is that the measured $x_{\textrm{sub}}$ of the samples of region-B arises mostly from the precipitates and a lack of Mn$_{\textrm{Si}}$ weakens the carrier mediated exchange interaction, which becomes dominated by a possible antiferromagnetic dipolar interaction between the precipitates. Another example of a low moment Mn-Si phase is amorphous Si$_{1-x}$Mn$_{x}$ alloys,\cite{Zeng:2008prb} where it has been suggested that the Mn moment is lost in an itinerant band created by Anderson localization.

\section{ The influence of the Pb surfactant }

The control samples grown without Pb demonstrate the dramatic influence of the Pb surfactant on the magnetic properties of the Mn $\delta$-doped layers. Without Pb, the samples in regions-A and C have no long-range magnetic order, as shown by the zero remanent magnetization of these samples in Fig.~\ref{fig:mrALL}(a) and (c), although the saturation magnetization for the control sample in region-A is 0.78 $\mu_B$.
Surprisingly, the situation is reversed in region B. The sample without Pb has a $T_C > 400$~K and a large remanent magnetization (Fig.~\ref{fig:mrALL}(b)). This behavior is explained in our previous work where a MnSi-B2 structure forms from the deposition of Mn on Si(001).\cite{Kahwaji:2012prb}  It is, however, worth pointing out that this B2 phase grown without Pb only forms at the higher coverages represented by region-B, and RBS indicates that the $x_{\textrm{sub}}$ is close to zero for this sample.
Like the other samples, we find that the $x_{\textrm{sub}}$ measured by RBS is not a good indicator of the magnetic behavior of the control samples.
The substitutional fraction for the control samples for regions-A and C shown in Fig.~\ref{fig:RBSRatios} is larger than the samples with a Pb-surfactant. It is likely that these Mn$_{sub}$ are also substitutional-like sites within the precipitates.  Support for this idea is found in the results of MBE grown Si$_{1-x}$Mn$_{x}$ alloys prepared by co-deposition of Mn and Si, where the alloy undergoes spinodal decomposition into nanocolumns.\cite{zhangJAP} For 1.5\% Mn, RBS shows a comparable $x_{sub}=0.7$ at $T_{\textrm{Si}} = 200~\celsius$. We have recently reproduced these growth conditions and also observed the formation of nanocolumns. However, despite the large $x_{\textrm{sub}}$ reported in Ref.~\onlinecite{zhangJAP}, these nanocolumns are paramagnetic with a small moment.\cite{Kahwaji:unpub}

\begin{figure}[h!]
\centering
\includegraphics [width=\columnwidth]{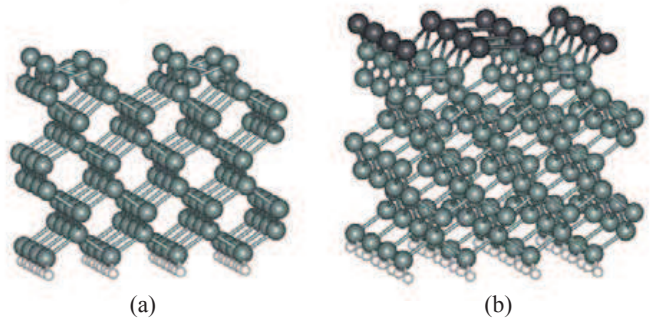}
 \caption{(Color online) (a) Schematic illustration of the simulation supercell. Middle (grey): Si. Bottom small (white): H. (b) Schematic illustration of the surface reconstruction of one monolayer of Pb on Si thin film. Top big (black): Pb. Middle: Si. Bottom small (white): H.}
 \label{fig:DFT}
\end{figure}

To understand how Pb can influence the formation of Mn$_{\textrm{Si}}$ we performed DFT calculations. In our study, we find that Pb increases the formation energy of Mn$_{\textrm{Si}}$ by direct substitution into a surface Si site.  However, the Pb surfactant tends to weaken the top layer bonding of Si because of the overly crowded bonding to the top Si atom.  Our DFT calculations show that this enhances vacancy formation in the top Si layers, and thereby creates a channel for the Mn to incorporate substitutionally. Various theoretical and experimental studies show that different amounts of Pb on Si can form different surface reconstructions \cite{{froyen}, {theosurf}, {expsurf}}. For one full monolayer coverage of Pb, we picked a $2\times1$ surface reconstruction as both confirmed by our \emph{in situ} RHEED observations and by scanning tunneling microscopy measurements and calculations \cite{{froyen}, {theosurf},{expsurf}}.
We performed DFT calculations using the VASP code within the generalized gradient approximation \cite{vasp1, vasp2}. We chose the Projected Augmented Wave PBE potential as our pseudo potential \cite{pbe1,pbe2}. Our simulation cell is a $144$ Si atoms slab with 32 H atoms passivating the dangling bonds of the bottom surface as shown in Fig.~\ref{fig:DFT}(a). One full layer of Pb atoms were introduced into the calculation as the surfactant. The thickness of the vacuum layer is about 18 {\AA}. We used a plane wave cutoff energy of $245$ eV and a $2\times2\times1$ k-point mesh for Brillouin zone sampling. A total energy minimization was performed by relaxing atomic positions until the forces converged to less than $0.01~eV/{\AA}$.  The calculated Si lattice constant is $5.469$ {\AA}, which is in good agreement with the experimental value of $5.43$ {\AA}.

We define the formation energy of a vacancy as $E(formation) = E(vac) - E(reference) + \mu(Si)$, where \emph{E(reference)} is the total energy of the Si slab, \emph{E(vac)} is the total energy of the same cell with one Si atom removed and $\mu(Si)$ is the chemical potential of the Si, which is the energy per atom of a bulk calculation of a system with 512 Si atoms. The chemical potential of the Si is calculated to be $-5.43$ eV.
We calculated the formation energy of vacancy of the atom on the surface layer in a pure Si thin film by removing a Si atom from the top dimer layer of the $2\times1$ reconstructed surface. We found that the vacancy formation energy in this case is $1.18$~eV. Next, we introduced Pb atoms as surfactants into the system, as shown in Fig.~\ref{fig:DFT} (b). We removed one Si atom from the top Si layer and calculated a vacancy formation energy of $0.26$~eV, which is about $0.92$~eV lower than the vacancy formation energy for a pure Si surface at the same site. This is mainly due to the Pb surfactant. The Pb atoms largely disturb the top Si surface layer and lower the Si vacancy formation energy. On the other hand, the interstitial diffusion is unlikely to change much due to the surfactant.

\section{Conclusion}

     Our results demonstrate the control of the incorporation of Mn $\delta$-doped layers into Si(001) via a Pb surfactant. We find that the resulting structure is sensitive to the growth conditions of the Si capping layer. Structural and magnetometry studies identify three distinct regions of the magnetic phase diagram, where two of them are ferromagnetic.  Samples in region-A of Fig.~\ref{fig:diagram} have a $T_C$ above 400~K and contain nanocrystalline MnSi precipitates with a B2-like crystal structure.  This high temperature ferromagnetism extends into region-C, where a second magnetic transition at $T_C = 170$~K exists due to MnSi nano-disks with a B20-like crystal structure.  This result complements previous work on MnSi$_{1.7}$ precipitates,\cite{Bolduc:2005prb,Affouda,ZhouPRB75,Orlov:2009jetp, Ko:2008jap} and provides examples of different nanocrystalline precipitates in Si.  More work needs to be done with MnSi$_{1.7}$ and MnSi-B20 to determine the role that magnetic defects play in these structures.\cite{Menshov:2011prb}

In this paper, we discuss a mechanism for an intriguing surfactant enhanced substitutional doping phenomena.  In the two ferromagnetic phases, RBS shows that an increase in $m_r$ is possibly correlated with an increase in substitutional-like Mn.  We interpret this increase as an increase in the concentration of substitutional Mn in the Si matrix, which may be the origin of $T_C > 400$~K phase.    DFT calculations show that a Pb layer on Si reduces the vacancy formation energy by $0.92$~eV which opens the pathway for substitutional doping of Mn into Si. This provides a new opportunity for doping in other DMS materials.

\begin{acknowledgements}
This work was supported by grants to EDC, MDR and TLM from the Natural Sciences and Engineering Research Council of Canada (NSERC). Work at Universit\'{e} de Montr\'{e}al was supported through grants from NSERC and FQRNT. XAFS experiments were carried out at the Advanced Photon Source (APS) using facilities of the PNC/XSD ID beamline, sector 20. Research at PNC/XCD facilities is supported by the US Department of Energy-Basic Energy Sciences (US DOE-BES), a MRS grant from NSERC, the University of Washington, Simon Fraser University and the APS. Use of the APS, an Office of Science User Facility operated for the U.S. DOE Office of Science by Argonne National Laboratory, was supported by the U.S. DOE under Contract No. DE-AC02-06CH11357. We thank S.-H. Wei for the useful discussions about DFT calculations.
\end{acknowledgements}


\begin{thebibliography}{57}%
\makeatletter
\providecommand \@ifxundefined [1]{%
 \@ifx{#1\undefined}
}%
\providecommand \@ifnum [1]{%
 \ifnum #1\expandafter \@firstoftwo
 \else \expandafter \@secondoftwo
 \fi
}%
\providecommand \@ifx [1]{%
 \ifx #1\expandafter \@firstoftwo
 \else \expandafter \@secondoftwo
 \fi
}%
\providecommand \natexlab [1]{#1}%
\providecommand \enquote  [1]{``#1''}%
\providecommand \bibnamefont  [1]{#1}%
\providecommand \bibfnamefont [1]{#1}%
\providecommand \citenamefont [1]{#1}%
\providecommand \href@noop [0]{\@secondoftwo}%
\providecommand \href [0]{\begingroup \@sanitize@url \@href}%
\providecommand \@href[1]{\@@startlink{#1}\@@href}%
\providecommand \@@href[1]{\endgroup#1\@@endlink}%
\providecommand \@sanitize@url [0]{\catcode `\\12\catcode `\$12\catcode
  `\&12\catcode `\#12\catcode `\^12\catcode `\_12\catcode `\%12\relax}%
\providecommand \@@startlink[1]{}%
\providecommand \@@endlink[0]{}%
\providecommand \url  [0]{\begingroup\@sanitize@url \@url }%
\providecommand \@url [1]{\endgroup\@href {#1}{\urlprefix }}%
\providecommand \urlprefix  [0]{URL }%
\providecommand \Eprint [0]{\href }%
\providecommand \doibase [0]{http://dx.doi.org/}%
\providecommand \selectlanguage [0]{\@gobble}%
\providecommand \bibinfo  [0]{\@secondoftwo}%
\providecommand \bibfield  [0]{\@secondoftwo}%
\providecommand \translation [1]{[#1]}%
\providecommand \BibitemOpen [0]{}%
\providecommand \bibitemStop [0]{}%
\providecommand \bibitemNoStop [0]{.\EOS\space}%
\providecommand \EOS [0]{\spacefactor3000\relax}%
\providecommand \BibitemShut  [1]{\csname bibitem#1\endcsname}%
\let\auto@bib@innerbib\@empty
\bibitem [{\citenamefont {Dietl}(2010)}]{Dietl:2010nm}%
  \BibitemOpen
  \bibfield  {author} {\bibinfo {author} {\bibfnamefont {T.}~\bibnamefont
  {Dietl}},\ }\href
  {http://www.nature.com/nmat/journal/v9/n12/full/nmat2898.html} {\bibfield
  {journal} {\bibinfo  {journal} {Nat. Mater.}\ }\textbf {\bibinfo {volume}
  {9}},\ \bibinfo {pages} {965 } (\bibinfo {year} {2010})}\BibitemShut
  {NoStop}%
\bibitem [{\citenamefont {Dietl}\ \emph {et~al.}(2000)\citenamefont {Dietl},
  \citenamefont {Ohno}, \citenamefont {Matsukura}, \citenamefont {Cibert},\
  and\ \citenamefont {Ferrand}}]{Dietl:2000sci}%
  \BibitemOpen
  \bibfield  {author} {\bibinfo {author} {\bibfnamefont {T.}~\bibnamefont
  {Dietl}}, \bibinfo {author} {\bibfnamefont {H.}~\bibnamefont {Ohno}},
  \bibinfo {author} {\bibfnamefont {F.}~\bibnamefont {Matsukura}}, \bibinfo
  {author} {\bibfnamefont {J.}~\bibnamefont {Cibert}}, \ and\ \bibinfo {author}
  {\bibfnamefont {D.}~\bibnamefont {Ferrand}},\ }\href {\doibase
  10.1126/science.287.5455.1019} {\bibfield  {journal} {\bibinfo  {journal}
  {Science}\ }\textbf {\bibinfo {volume} {287}},\ \bibinfo {pages} {1019}
  (\bibinfo {year} {2000})}\BibitemShut {NoStop}%
\bibitem [{\citenamefont {Stroppa}\ \emph {et~al.}(2003)\citenamefont
  {Stroppa}, \citenamefont {Picozzi}, \citenamefont {Continenza},\ and\
  \citenamefont {Freeman}}]{Stroppa:2003prb}%
  \BibitemOpen
  \bibfield  {author} {\bibinfo {author} {\bibfnamefont {A.}~\bibnamefont
  {Stroppa}}, \bibinfo {author} {\bibfnamefont {S.}~\bibnamefont {Picozzi}},
  \bibinfo {author} {\bibfnamefont {A.}~\bibnamefont {Continenza}}, \ and\
  \bibinfo {author} {\bibfnamefont {A.~J.}\ \bibnamefont {Freeman}},\ }\href
  {\doibase 10.1103/PhysRevB.68.155203} {\bibfield  {journal} {\bibinfo
  {journal} {Phys. Rev. B}\ }\textbf {\bibinfo {volume} {68}},\ \bibinfo
  {pages} {155203} (\bibinfo {year} {2003})}\BibitemShut {NoStop}%
\bibitem [{\citenamefont {Eaglesham}\ \emph {et~al.}(1990)\citenamefont
  {Eaglesham}, \citenamefont {Gossmann},\ and\ \citenamefont
  {Cerullo}}]{Eaglesham:1990prl}%
  \BibitemOpen
  \bibfield  {author} {\bibinfo {author} {\bibfnamefont {D.~J.}\ \bibnamefont
  {Eaglesham}}, \bibinfo {author} {\bibfnamefont {H.-J.}\ \bibnamefont
  {Gossmann}}, \ and\ \bibinfo {author} {\bibfnamefont {M.}~\bibnamefont
  {Cerullo}},\ }\href {\doibase 10.1103/PhysRevLett.65.1227} {\bibfield
  {journal} {\bibinfo  {journal} {Phys. Rev. Lett.}\ }\textbf {\bibinfo
  {volume} {65}},\ \bibinfo {pages} {1227} (\bibinfo {year}
  {1990})}\BibitemShut {NoStop}%
\bibitem [{\citenamefont {Zhang}\ \emph {et~al.}(2005)\citenamefont {Zhang},
  \citenamefont {Jiang}, \citenamefont {Smith},\ and\ \citenamefont
  {Drucker}}]{zhangJAP}%
  \BibitemOpen
  \bibfield  {author} {\bibinfo {author} {\bibfnamefont {Y.}~\bibnamefont
  {Zhang}}, \bibinfo {author} {\bibfnamefont {Q.}~\bibnamefont {Jiang}},
  \bibinfo {author} {\bibfnamefont {D.~J.}\ \bibnamefont {Smith}}, \ and\
  \bibinfo {author} {\bibfnamefont {J.}~\bibnamefont {Drucker}},\ }\href
  {\doibase 10.1063/1.1988973} {\bibfield  {journal} {\bibinfo  {journal} {J.
  Appl. Phys.}\ }\textbf {\bibinfo {volume} {98}},\ \bibinfo {eid} {033512}
  (\bibinfo {year} {2005})}\BibitemShut {NoStop}%
\bibitem [{\citenamefont {Awo-Affouda}\ \emph {et~al.}(2006)\citenamefont
  {Awo-Affouda}, \citenamefont {Bolduc}, \citenamefont {Huang}, \citenamefont
  {Ramos}, \citenamefont {Dunn}, \citenamefont {Thiel}, \citenamefont
  {Agnello},\ and\ \citenamefont {LaBella}}]{Affouda}%
  \BibitemOpen
  \bibfield  {author} {\bibinfo {author} {\bibfnamefont {C.}~\bibnamefont
  {Awo-Affouda}}, \bibinfo {author} {\bibfnamefont {M.}~\bibnamefont {Bolduc}},
  \bibinfo {author} {\bibfnamefont {M.~B.}\ \bibnamefont {Huang}}, \bibinfo
  {author} {\bibfnamefont {F.~G.}\ \bibnamefont {Ramos}}, \bibinfo {author}
  {\bibfnamefont {K.~A.}\ \bibnamefont {Dunn}}, \bibinfo {author}
  {\bibfnamefont {B.}~\bibnamefont {Thiel}}, \bibinfo {author} {\bibfnamefont
  {G.}~\bibnamefont {Agnello}}, \ and\ \bibinfo {author} {\bibfnamefont
  {V.~P.}\ \bibnamefont {LaBella}},\ }\href@noop {} {\bibfield  {journal}
  {\bibinfo  {journal} {J. Vac. Sci. Technol. A}\ }\textbf {\bibinfo {volume}
  {24}},\ \bibinfo {pages} {1644} (\bibinfo {year} {2006})}\BibitemShut
  {NoStop}%
\bibitem [{\citenamefont {Zhang}\ \emph {et~al.}(2004)\citenamefont {Zhang},
  \citenamefont {Liu}, \citenamefont {Gao}, \citenamefont {Wu}, \citenamefont
  {Du}, \citenamefont {Zhu}, \citenamefont {Xiao},\ and\ \citenamefont
  {Chen}}]{Zhang}%
  \BibitemOpen
  \bibfield  {author} {\bibinfo {author} {\bibfnamefont {F.~M.}\ \bibnamefont
  {Zhang}}, \bibinfo {author} {\bibfnamefont {X.~C.}\ \bibnamefont {Liu}},
  \bibinfo {author} {\bibfnamefont {J.}~\bibnamefont {Gao}}, \bibinfo {author}
  {\bibfnamefont {X.~S.}\ \bibnamefont {Wu}}, \bibinfo {author} {\bibfnamefont
  {Y.~W.}\ \bibnamefont {Du}}, \bibinfo {author} {\bibfnamefont
  {H.}~\bibnamefont {Zhu}}, \bibinfo {author} {\bibfnamefont {J.~Q.}\
  \bibnamefont {Xiao}}, \ and\ \bibinfo {author} {\bibfnamefont
  {P.}~\bibnamefont {Chen}},\ }\href {http://dx.doi.org/doi/10.1063/1.1775886}
  {\bibfield  {journal} {\bibinfo  {journal} {Appl. Phys. Lett.}\ }\textbf
  {\bibinfo {volume} {85}},\ \bibinfo {pages} {786} (\bibinfo {year}
  {2004})}\BibitemShut {NoStop}%
\bibitem [{\citenamefont {Bolduc}\ \emph {et~al.}(2005)\citenamefont {Bolduc},
  \citenamefont {Awo-Affouda}, \citenamefont {Stollenwerk}, \citenamefont
  {Huang}, \citenamefont {Ramos}, \citenamefont {Agnello},\ and\ \citenamefont
  {LaBella}}]{Bolduc:2005prb}%
  \BibitemOpen
  \bibfield  {author} {\bibinfo {author} {\bibfnamefont {M.}~\bibnamefont
  {Bolduc}}, \bibinfo {author} {\bibfnamefont {C.}~\bibnamefont {Awo-Affouda}},
  \bibinfo {author} {\bibfnamefont {A.}~\bibnamefont {Stollenwerk}}, \bibinfo
  {author} {\bibfnamefont {M.~B.}\ \bibnamefont {Huang}}, \bibinfo {author}
  {\bibfnamefont {F.~G.}\ \bibnamefont {Ramos}}, \bibinfo {author}
  {\bibfnamefont {G.}~\bibnamefont {Agnello}}, \ and\ \bibinfo {author}
  {\bibfnamefont {V.~P.}\ \bibnamefont {LaBella}},\ }\href
  {http://link.aps.org/abstract/PRB/v71/e033302} {\bibfield  {journal}
  {\bibinfo  {journal} {Phys. Rev. B}\ }\textbf {\bibinfo {volume} {71}},\
  \bibinfo {pages} {033302} (\bibinfo {year} {2005})}\BibitemShut {NoStop}%
\bibitem [{\citenamefont {Ko}\ \emph {et~al.}(2008)\citenamefont {Ko},
  \citenamefont {Teo}, \citenamefont {Liew}, \citenamefont {Chong},
  \citenamefont {MacKenzie}, \citenamefont {MacLaren},\ and\ \citenamefont
  {Chapman}}]{Ko:2008jap}%
  \BibitemOpen
  \bibfield  {author} {\bibinfo {author} {\bibfnamefont {V.}~\bibnamefont
  {Ko}}, \bibinfo {author} {\bibfnamefont {K.~L.}\ \bibnamefont {Teo}},
  \bibinfo {author} {\bibfnamefont {T.}~\bibnamefont {Liew}}, \bibinfo {author}
  {\bibfnamefont {T.~C.}\ \bibnamefont {Chong}}, \bibinfo {author}
  {\bibfnamefont {M.}~\bibnamefont {MacKenzie}}, \bibinfo {author}
  {\bibfnamefont {I.}~\bibnamefont {MacLaren}}, \ and\ \bibinfo {author}
  {\bibfnamefont {J.~N.}\ \bibnamefont {Chapman}},\ }\href
  {http://link.aip.org/link/?JAP/104/033912/1} {\bibfield  {journal} {\bibinfo
  {journal} {J. Appl. Phys.}\ }\textbf {\bibinfo {volume} {104}},\ \bibinfo
  {pages} {033912} (\bibinfo {year} {2008})}\BibitemShut {NoStop}%
\bibitem [{\citenamefont {Men'shov}\ \emph {et~al.}(2011)\citenamefont
  {Men'shov}, \citenamefont {Tugushev}, \citenamefont {Caprara},\ and\
  \citenamefont {Chulkov}}]{Menshov:2011prb}%
  \BibitemOpen
  \bibfield  {author} {\bibinfo {author} {\bibfnamefont {V.~N.}\ \bibnamefont
  {Men'shov}}, \bibinfo {author} {\bibfnamefont {V.~V.}\ \bibnamefont
  {Tugushev}}, \bibinfo {author} {\bibfnamefont {S.}~\bibnamefont {Caprara}}, \
  and\ \bibinfo {author} {\bibfnamefont {E.~V.}\ \bibnamefont {Chulkov}},\
  }\href {\doibase 10.1103/PhysRevB.83.035201} {\bibfield  {journal} {\bibinfo
  {journal} {Phys. Rev. B}\ }\textbf {\bibinfo {volume} {83}},\ \bibinfo
  {pages} {035201} (\bibinfo {year} {2011})}\BibitemShut {NoStop}%
\bibitem [{\citenamefont {Zhou}\ \emph {et~al.}(2007)\citenamefont {Zhou},
  \citenamefont {Potzger}, \citenamefont {Zhang}, \citenamefont {M\"ucklich},
  \citenamefont {Eichhorn}, \citenamefont {Schell}, \citenamefont
  {Gr\"otzschel}, \citenamefont {Schmidt}, \citenamefont {Skorupa},
  \citenamefont {Helm}, \citenamefont {Fassbender},\ and\ \citenamefont
  {Geiger}}]{ZhouPRB75}%
  \BibitemOpen
  \bibfield  {author} {\bibinfo {author} {\bibfnamefont {S.}~\bibnamefont
  {Zhou}}, \bibinfo {author} {\bibfnamefont {K.}~\bibnamefont {Potzger}},
  \bibinfo {author} {\bibfnamefont {G.}~\bibnamefont {Zhang}}, \bibinfo
  {author} {\bibfnamefont {A.}~\bibnamefont {M\"ucklich}}, \bibinfo {author}
  {\bibfnamefont {F.}~\bibnamefont {Eichhorn}}, \bibinfo {author}
  {\bibfnamefont {N.}~\bibnamefont {Schell}}, \bibinfo {author} {\bibfnamefont
  {R.}~\bibnamefont {Gr\"otzschel}}, \bibinfo {author} {\bibfnamefont
  {B.}~\bibnamefont {Schmidt}}, \bibinfo {author} {\bibfnamefont
  {W.}~\bibnamefont {Skorupa}}, \bibinfo {author} {\bibfnamefont
  {M.}~\bibnamefont {Helm}}, \bibinfo {author} {\bibfnamefont {J.}~\bibnamefont
  {Fassbender}}, \ and\ \bibinfo {author} {\bibfnamefont {D.}~\bibnamefont
  {Geiger}},\ }\href {\doibase 10.1103/PhysRevB.75.085203} {\bibfield
  {journal} {\bibinfo  {journal} {Phys. Rev. B}\ }\textbf {\bibinfo {volume}
  {75}},\ \bibinfo {pages} {085203} (\bibinfo {year} {2007})}\BibitemShut
  {NoStop}%
\bibitem [{\citenamefont {Orlov}\ \emph {et~al.}(2009)\citenamefont {Orlov},
  \citenamefont {Granovsky}, \citenamefont {Balagurov}, \citenamefont
  {Kulemanov}, \citenamefont {Parkhomenko}, \citenamefont {Perov},
  \citenamefont {Gan'shina}, \citenamefont {Bublik}, \citenamefont
  {Shcherbachev}, \citenamefont {Kartavykh}, \citenamefont {Vdovin},
  \citenamefont {Sapelkin}, \citenamefont {Saraikin}, \citenamefont {Agafonov},
  \citenamefont {Zinenko}, \citenamefont {Rogalev},\ and\ \citenamefont
  {Smekhova}}]{Orlov:2009jetp}%
  \BibitemOpen
  \bibfield  {author} {\bibinfo {author} {\bibfnamefont {A.}~\bibnamefont
  {Orlov}}, \bibinfo {author} {\bibfnamefont {A.}~\bibnamefont {Granovsky}},
  \bibinfo {author} {\bibfnamefont {L.}~\bibnamefont {Balagurov}}, \bibinfo
  {author} {\bibfnamefont {I.}~\bibnamefont {Kulemanov}}, \bibinfo {author}
  {\bibfnamefont {Y.}~\bibnamefont {Parkhomenko}}, \bibinfo {author}
  {\bibfnamefont {N.}~\bibnamefont {Perov}}, \bibinfo {author} {\bibfnamefont
  {E.}~\bibnamefont {Gan'shina}}, \bibinfo {author} {\bibfnamefont
  {V.}~\bibnamefont {Bublik}}, \bibinfo {author} {\bibfnamefont
  {K.}~\bibnamefont {Shcherbachev}}, \bibinfo {author} {\bibfnamefont
  {A.}~\bibnamefont {Kartavykh}}, \bibinfo {author} {\bibfnamefont
  {V.}~\bibnamefont {Vdovin}}, \bibinfo {author} {\bibfnamefont
  {A.}~\bibnamefont {Sapelkin}}, \bibinfo {author} {\bibfnamefont
  {V.}~\bibnamefont {Saraikin}}, \bibinfo {author} {\bibfnamefont
  {Y.}~\bibnamefont {Agafonov}}, \bibinfo {author} {\bibfnamefont
  {V.}~\bibnamefont {Zinenko}}, \bibinfo {author} {\bibfnamefont
  {A.}~\bibnamefont {Rogalev}}, \ and\ \bibinfo {author} {\bibfnamefont
  {A.}~\bibnamefont {Smekhova}},\ }\href
  {http://dx.doi.org/10.1134/S1063776109100069} {\bibfield  {journal} {\bibinfo
   {journal} {J. Exp. Theor. Phys.}\ }\textbf {\bibinfo {volume} {109}},\
  \bibinfo {pages} {602} (\bibinfo {year} {2009})}\BibitemShut {NoStop}%
\bibitem [{\citenamefont {Kawakami}\ \emph {et~al.}(2000)\citenamefont
  {Kawakami}, \citenamefont {Johnston-Halperin}, \citenamefont {Chen},
  \citenamefont {Hanson}, \citenamefont {Gu\'{e}bels}, \citenamefont {Speck},
  \citenamefont {Gossard},\ and\ \citenamefont {Awschalom}}]{Kawakami:2000apl}%
  \BibitemOpen
  \bibfield  {author} {\bibinfo {author} {\bibfnamefont {R.~K.}\ \bibnamefont
  {Kawakami}}, \bibinfo {author} {\bibfnamefont {E.}~\bibnamefont
  {Johnston-Halperin}}, \bibinfo {author} {\bibfnamefont {L.~F.}\ \bibnamefont
  {Chen}}, \bibinfo {author} {\bibfnamefont {M.}~\bibnamefont {Hanson}},
  \bibinfo {author} {\bibfnamefont {N.}~\bibnamefont {Gu\'{e}bels}}, \bibinfo
  {author} {\bibfnamefont {J.~S.}\ \bibnamefont {Speck}}, \bibinfo {author}
  {\bibfnamefont {A.~C.}\ \bibnamefont {Gossard}}, \ and\ \bibinfo {author}
  {\bibfnamefont {D.~D.}\ \bibnamefont {Awschalom}},\ }\href {\doibase
  10.1063/1.1316775} {\bibfield  {journal} {\bibinfo  {journal} {Appl. Phys.
  Lett.}\ }\textbf {\bibinfo {volume} {77}},\ \bibinfo {pages} {2379} (\bibinfo
  {year} {2000})}\BibitemShut {NoStop}%
\bibitem [{\citenamefont {Chen}\ \emph {et~al.}(2002)\citenamefont {Chen},
  \citenamefont {Na}, \citenamefont {Cheon}, \citenamefont {Wang},
  \citenamefont {Luo}, \citenamefont {McCombe}, \citenamefont {Liu},
  \citenamefont {Sasaki}, \citenamefont {Wojtowicz}, \citenamefont {Furdyna},
  \citenamefont {Potashnik},\ and\ \citenamefont {Schiffer}}]{Chen:2002apl}%
  \BibitemOpen
  \bibfield  {author} {\bibinfo {author} {\bibfnamefont {X.}~\bibnamefont
  {Chen}}, \bibinfo {author} {\bibfnamefont {M.}~\bibnamefont {Na}}, \bibinfo
  {author} {\bibfnamefont {M.}~\bibnamefont {Cheon}}, \bibinfo {author}
  {\bibfnamefont {S.}~\bibnamefont {Wang}}, \bibinfo {author} {\bibfnamefont
  {H.}~\bibnamefont {Luo}}, \bibinfo {author} {\bibfnamefont {B.~D.}\
  \bibnamefont {McCombe}}, \bibinfo {author} {\bibfnamefont {X.}~\bibnamefont
  {Liu}}, \bibinfo {author} {\bibfnamefont {Y.}~\bibnamefont {Sasaki}},
  \bibinfo {author} {\bibfnamefont {T.}~\bibnamefont {Wojtowicz}}, \bibinfo
  {author} {\bibfnamefont {J.~K.}\ \bibnamefont {Furdyna}}, \bibinfo {author}
  {\bibfnamefont {S.~J.}\ \bibnamefont {Potashnik}}, \ and\ \bibinfo {author}
  {\bibfnamefont {P.}~\bibnamefont {Schiffer}},\ }\href {\doibase
  10.1063/1.1481184} {\bibfield  {journal} {\bibinfo  {journal} {Appl. Phys.
  Lett.}\ }\textbf {\bibinfo {volume} {81}},\ \bibinfo {pages} {511} (\bibinfo
  {year} {2002})}\BibitemShut {NoStop}%
\bibitem [{\citenamefont {Qian}\ \emph {et~al.}(2006)\citenamefont {Qian},
  \citenamefont {Fong}, \citenamefont {Liu}, \citenamefont {Pickett},
  \citenamefont {Pask},\ and\ \citenamefont {Yang}}]{Qian:2006prl}%
  \BibitemOpen
  \bibfield  {author} {\bibinfo {author} {\bibfnamefont {M.~C.}\ \bibnamefont
  {Qian}}, \bibinfo {author} {\bibfnamefont {C.~Y.}\ \bibnamefont {Fong}},
  \bibinfo {author} {\bibfnamefont {K.}~\bibnamefont {Liu}}, \bibinfo {author}
  {\bibfnamefont {W.~E.}\ \bibnamefont {Pickett}}, \bibinfo {author}
  {\bibfnamefont {J.~E.}\ \bibnamefont {Pask}}, \ and\ \bibinfo {author}
  {\bibfnamefont {L.~H.}\ \bibnamefont {Yang}},\ }\href {\doibase
  10.1103/PhysRevLett.96.027211} {\bibfield  {journal} {\bibinfo  {journal}
  {Phys. Rev. Lett.}\ }\textbf {\bibinfo {volume} {96}},\ \bibinfo {pages}
  {027211} (\bibinfo {year} {2006})}\BibitemShut {NoStop}%
\bibitem [{\citenamefont {Krause}\ \emph {et~al.}(2007)\citenamefont {Krause},
  \citenamefont {Stollenwerk}, \citenamefont {Reed}, \citenamefont {LaBella},
  \citenamefont {Hortamani}, \citenamefont {Kratzer},\ and\ \citenamefont
  {Scheffler}}]{Krause:2007prb}%
  \BibitemOpen
  \bibfield  {author} {\bibinfo {author} {\bibfnamefont {M.~R.}\ \bibnamefont
  {Krause}}, \bibinfo {author} {\bibfnamefont {A.~J.}\ \bibnamefont
  {Stollenwerk}}, \bibinfo {author} {\bibfnamefont {J.}~\bibnamefont {Reed}},
  \bibinfo {author} {\bibfnamefont {V.~P.}\ \bibnamefont {LaBella}}, \bibinfo
  {author} {\bibfnamefont {M.}~\bibnamefont {Hortamani}}, \bibinfo {author}
  {\bibfnamefont {P.}~\bibnamefont {Kratzer}}, \ and\ \bibinfo {author}
  {\bibfnamefont {M.}~\bibnamefont {Scheffler}},\ }\href {\doibase
  10.1103/PhysRevB.75.205326} {\bibfield  {journal} {\bibinfo  {journal} {Phys.
  Rev. B}\ }\textbf {\bibinfo {volume} {75}},\ \bibinfo {pages} {205326}
  (\bibinfo {year} {2007})}\BibitemShut {NoStop}%
\bibitem [{\citenamefont {Liu}\ and\ \citenamefont
  {Reinke}(2008)}]{Liu:2008ss}%
  \BibitemOpen
  \bibfield  {author} {\bibinfo {author} {\bibfnamefont {H.}~\bibnamefont
  {Liu}}\ and\ \bibinfo {author} {\bibfnamefont {P.}~\bibnamefont {Reinke}},\
  }\href {\doibase 10.1016/j.susc.2007.12.043} {\bibfield  {journal} {\bibinfo
  {journal} {Surf. Sci.}\ }\textbf {\bibinfo {volume} {602}},\ \bibinfo {pages}
  {986 } (\bibinfo {year} {2008})}\BibitemShut {NoStop}%
\bibitem [{\citenamefont {Nolph}\ \emph {et~al.}(2011)\citenamefont {Nolph},
  \citenamefont {Liu},\ and\ \citenamefont {Reinke}}]{Nolph:2011ss}%
  \BibitemOpen
  \bibfield  {author} {\bibinfo {author} {\bibfnamefont {C.}~\bibnamefont
  {Nolph}}, \bibinfo {author} {\bibfnamefont {H.}~\bibnamefont {Liu}}, \ and\
  \bibinfo {author} {\bibfnamefont {P.}~\bibnamefont {Reinke}},\ }\href
  {\doibase DOI: 10.1016/j.susc.2011.04.017} {\bibfield  {journal} {\bibinfo
  {journal} {Surf. Sci.}\ }\textbf {\bibinfo {volume} {605}},\ \bibinfo {pages}
  {L29 } (\bibinfo {year} {2011})}\BibitemShut {NoStop}%
\bibitem [{\citenamefont {Wu}\ \emph {et~al.}(2007)\citenamefont {Wu},
  \citenamefont {Kratzer},\ and\ \citenamefont {Scheffler}}]{Wu:2007prl}%
  \BibitemOpen
  \bibfield  {author} {\bibinfo {author} {\bibfnamefont {H.}~\bibnamefont
  {Wu}}, \bibinfo {author} {\bibfnamefont {P.}~\bibnamefont {Kratzer}}, \ and\
  \bibinfo {author} {\bibfnamefont {M.}~\bibnamefont {Scheffler}},\ }\href
  {\doibase 10.1103/PhysRevLett.98.117202} {\bibfield  {journal} {\bibinfo
  {journal} {Phys. Rev. Lett.}\ }\textbf {\bibinfo {volume} {98}},\ \bibinfo
  {pages} {117202} (\bibinfo {year} {2007})}\BibitemShut {NoStop}%
\bibitem [{\citenamefont {Bernardini}\ \emph {et~al.}(2004)\citenamefont
  {Bernardini}, \citenamefont {Picozzi},\ and\ \citenamefont
  {Continenza}}]{Bernardini:2004apl}%
  \BibitemOpen
  \bibfield  {author} {\bibinfo {author} {\bibfnamefont {F.}~\bibnamefont
  {Bernardini}}, \bibinfo {author} {\bibfnamefont {S.}~\bibnamefont {Picozzi}},
  \ and\ \bibinfo {author} {\bibfnamefont {A.}~\bibnamefont {Continenza}},\
  }\href {\doibase 10.1063/1.1688002} {\bibfield  {journal} {\bibinfo
  {journal} {Appl. Phys. Lett.}\ }\textbf {\bibinfo {volume} {84}},\ \bibinfo
  {pages} {2289} (\bibinfo {year} {2004})}\BibitemShut {NoStop}%
\bibitem [{\citenamefont {Liu}\ \emph {et~al.}(2008)\citenamefont {Liu},
  \citenamefont {Yan}, \citenamefont {Wei}, \citenamefont {Sun}, \citenamefont
  {Pan}, \citenamefont {Soldatov}, \citenamefont {Mai}, \citenamefont {Pei},
  \citenamefont {Zhang}, \citenamefont {Jiang},\ and\ \citenamefont
  {Wei}}]{Liu:2008prb}%
  \BibitemOpen
  \bibfield  {author} {\bibinfo {author} {\bibfnamefont {Q.}~\bibnamefont
  {Liu}}, \bibinfo {author} {\bibfnamefont {W.}~\bibnamefont {Yan}}, \bibinfo
  {author} {\bibfnamefont {H.}~\bibnamefont {Wei}}, \bibinfo {author}
  {\bibfnamefont {Z.}~\bibnamefont {Sun}}, \bibinfo {author} {\bibfnamefont
  {Z.}~\bibnamefont {Pan}}, \bibinfo {author} {\bibfnamefont {A.~V.}\
  \bibnamefont {Soldatov}}, \bibinfo {author} {\bibfnamefont {C.}~\bibnamefont
  {Mai}}, \bibinfo {author} {\bibfnamefont {C.}~\bibnamefont {Pei}}, \bibinfo
  {author} {\bibfnamefont {X.}~\bibnamefont {Zhang}}, \bibinfo {author}
  {\bibfnamefont {Y.}~\bibnamefont {Jiang}}, \ and\ \bibinfo {author}
  {\bibfnamefont {S.}~\bibnamefont {Wei}},\ }\href {\doibase
  10.1103/PhysRevB.77.245211} {\bibfield  {journal} {\bibinfo  {journal} {Phys.
  Rev. B}\ }\textbf {\bibinfo {volume} {77}},\ \bibinfo {pages} {245211}
  (\bibinfo {year} {2008})}\BibitemShut {NoStop}%
\bibitem [{\citenamefont {Otrokov}\ \emph {et~al.}(2011)\citenamefont
  {Otrokov}, \citenamefont {Ernst}, \citenamefont {Tugushev}, \citenamefont
  {Ostanin}, \citenamefont {Buczek}, \citenamefont {Sandratskii}, \citenamefont
  {Fischer}, \citenamefont {Hergert}, \citenamefont {Mertig}, \citenamefont
  {Kuznetsov},\ and\ \citenamefont {Chulkov}}]{Otrokov:2011prb}%
  \BibitemOpen
  \bibfield  {author} {\bibinfo {author} {\bibfnamefont {M.~M.}\ \bibnamefont
  {Otrokov}}, \bibinfo {author} {\bibfnamefont {A.}~\bibnamefont {Ernst}},
  \bibinfo {author} {\bibfnamefont {V.~V.}\ \bibnamefont {Tugushev}}, \bibinfo
  {author} {\bibfnamefont {S.}~\bibnamefont {Ostanin}}, \bibinfo {author}
  {\bibfnamefont {P.}~\bibnamefont {Buczek}}, \bibinfo {author} {\bibfnamefont
  {L.~M.}\ \bibnamefont {Sandratskii}}, \bibinfo {author} {\bibfnamefont
  {G.}~\bibnamefont {Fischer}}, \bibinfo {author} {\bibfnamefont
  {W.}~\bibnamefont {Hergert}}, \bibinfo {author} {\bibfnamefont
  {I.}~\bibnamefont {Mertig}}, \bibinfo {author} {\bibfnamefont {V.~M.}\
  \bibnamefont {Kuznetsov}}, \ and\ \bibinfo {author} {\bibfnamefont {E.~V.}\
  \bibnamefont {Chulkov}},\ }\href {\doibase 10.1103/PhysRevB.84.144431}
  {\bibfield  {journal} {\bibinfo  {journal} {Phys. Rev. B}\ }\textbf {\bibinfo
  {volume} {84}},\ \bibinfo {pages} {144431} (\bibinfo {year}
  {2011})}\BibitemShut {NoStop}%
\bibitem [{\citenamefont {Krause}\ \emph {et~al.}(2006)\citenamefont {Krause},
  \citenamefont {Stollenwerk}, \citenamefont {Licurse},\ and\ \citenamefont
  {LaBella}}]{Krause:2006jvsta}%
  \BibitemOpen
  \bibfield  {author} {\bibinfo {author} {\bibfnamefont {M.~R.}\ \bibnamefont
  {Krause}}, \bibinfo {author} {\bibfnamefont {A.}~\bibnamefont {Stollenwerk}},
  \bibinfo {author} {\bibfnamefont {M.}~\bibnamefont {Licurse}}, \ and\
  \bibinfo {author} {\bibfnamefont {V.~P.}\ \bibnamefont {LaBella}},\ }\href
  {\doibase 10.1116/1.2167070} {\bibfield  {journal} {\bibinfo  {journal} {J.
  Vac. Sci. Technol.}\ }\textbf {\bibinfo {volume} {24}},\ \bibinfo {pages}
  {1480} (\bibinfo {year} {2006})}\BibitemShut {NoStop}%
\bibitem [{\citenamefont {Voigtl{\"a}nder}\ \emph {et~al.}(1995)\citenamefont
  {Voigtl{\"a}nder}, \citenamefont {Zinner}, \citenamefont {Weber},\ and\
  \citenamefont {Bonzel}}]{Voigtlander:1995prb}%
  \BibitemOpen
  \bibfield  {author} {\bibinfo {author} {\bibfnamefont {B.}~\bibnamefont
  {Voigtl{\"a}nder}}, \bibinfo {author} {\bibfnamefont {A.}~\bibnamefont
  {Zinner}}, \bibinfo {author} {\bibfnamefont {T.}~\bibnamefont {Weber}}, \
  and\ \bibinfo {author} {\bibfnamefont {H.~P.}\ \bibnamefont {Bonzel}},\
  }\href@noop {} {\bibfield  {journal} {\bibinfo  {journal} {Phys. Rev. B}\
  }\textbf {\bibinfo {volume} {51}},\ \bibinfo {pages} {7583} (\bibinfo {year}
  {1995})}\BibitemShut {NoStop}%
\bibitem [{\citenamefont {Evans}\ \emph {et~al.}(1996)\citenamefont {Evans},
  \citenamefont {Glueckstein},\ and\ \citenamefont {Nogami}}]{Evans:1996prb}%
  \BibitemOpen
  \bibfield  {author} {\bibinfo {author} {\bibfnamefont {M.~M.~R.}\
  \bibnamefont {Evans}}, \bibinfo {author} {\bibfnamefont {J.~C.}\ \bibnamefont
  {Glueckstein}}, \ and\ \bibinfo {author} {\bibfnamefont {J.}~\bibnamefont
  {Nogami}},\ }\href {\doibase FEB 15} {\bibfield  {journal} {\bibinfo
  {journal} {Phys. Rev. B}\ }\textbf {\bibinfo {volume} {53}},\ \bibinfo
  {pages} {4000} (\bibinfo {year} {1996})}\BibitemShut {NoStop}%
\bibitem [{\citenamefont {Dubon}\ \emph {et~al.}(2001)\citenamefont {Dubon},
  \citenamefont {Evans}, \citenamefont {Chervinsky}, \citenamefont {Aziz},
  \citenamefont {Spaepen}, \citenamefont {Golovchenko}, \citenamefont
  {Chisholm},\ and\ \citenamefont {Muller}}]{Dubon:2001apl}%
  \BibitemOpen
  \bibfield  {author} {\bibinfo {author} {\bibfnamefont {O.~D.}\ \bibnamefont
  {Dubon}}, \bibinfo {author} {\bibfnamefont {P.~G.}\ \bibnamefont {Evans}},
  \bibinfo {author} {\bibfnamefont {J.~F.}\ \bibnamefont {Chervinsky}},
  \bibinfo {author} {\bibfnamefont {M.~J.}\ \bibnamefont {Aziz}}, \bibinfo
  {author} {\bibfnamefont {F.}~\bibnamefont {Spaepen}}, \bibinfo {author}
  {\bibfnamefont {J.~A.}\ \bibnamefont {Golovchenko}}, \bibinfo {author}
  {\bibfnamefont {M.~F.}\ \bibnamefont {Chisholm}}, \ and\ \bibinfo {author}
  {\bibfnamefont {D.~A.}\ \bibnamefont {Muller}},\ }\href
  {http://link.aip.org/link/?APL/78/1505/1} {\bibfield  {journal} {\bibinfo
  {journal} {Appl. Phys. Lett.}\ }\textbf {\bibinfo {volume} {78}},\ \bibinfo
  {pages} {1505} (\bibinfo {year} {2001})}\BibitemShut {NoStop}%
\bibitem [{\citenamefont {Zhu}\ \emph {et~al.}(2008)\citenamefont {Zhu},
  \citenamefont {Liu},\ and\ \citenamefont {Stringfellow}}]{zhuprl}%
  \BibitemOpen
  \bibfield  {author} {\bibinfo {author} {\bibfnamefont {J.~Y.}\ \bibnamefont
  {Zhu}}, \bibinfo {author} {\bibfnamefont {F.}~\bibnamefont {Liu}}, \ and\
  \bibinfo {author} {\bibfnamefont {G.~B.}\ \bibnamefont {Stringfellow}},\
  }\href {\doibase 10.1103/PhysRevLett.101.196103} {\bibfield  {journal}
  {\bibinfo  {journal} {Phys. Rev. Lett.}\ }\textbf {\bibinfo {volume} {101}},\
  \bibinfo {pages} {196103} (\bibinfo {year} {2008})}\BibitemShut {NoStop}%
\bibitem [{\citenamefont {Zhu}\ \emph {et~al.}(2010)\citenamefont {Zhu},
  \citenamefont {Liu},\ and\ \citenamefont {Stringfellow}}]{zhujcg}%
  \BibitemOpen
  \bibfield  {author} {\bibinfo {author} {\bibfnamefont {J.}~\bibnamefont
  {Zhu}}, \bibinfo {author} {\bibfnamefont {F.}~\bibnamefont {Liu}}, \ and\
  \bibinfo {author} {\bibfnamefont {G.}~\bibnamefont {Stringfellow}},\ }\href
  {\doibase 10.1016/j.jcrysgro.2009.10.031} {\bibfield  {journal} {\bibinfo
  {journal} {J. Cryst. Growth}\ }\textbf {\bibinfo {volume} {312}},\ \bibinfo
  {pages} {174 } (\bibinfo {year} {2010})}\BibitemShut {NoStop}%
\bibitem [{\citenamefont {Zhang}\ \emph {et~al.}(2009)\citenamefont {Zhang},
  \citenamefont {Yan},\ and\ \citenamefont {Wei}}]{lixin}%
  \BibitemOpen
  \bibfield  {author} {\bibinfo {author} {\bibfnamefont {L.}~\bibnamefont
  {Zhang}}, \bibinfo {author} {\bibfnamefont {Y.}~\bibnamefont {Yan}}, \ and\
  \bibinfo {author} {\bibfnamefont {S.-H.}\ \bibnamefont {Wei}},\ }\href
  {\doibase 10.1103/PhysRevB.80.073305} {\bibfield  {journal} {\bibinfo
  {journal} {Phys. Rev. B}\ }\textbf {\bibinfo {volume} {80}},\ \bibinfo
  {pages} {073305} (\bibinfo {year} {2009})}\BibitemShut {NoStop}%
\bibitem [{\citenamefont {Xiao}\ \emph {et~al.}(2009)\citenamefont {Xiao},
  \citenamefont {Kahwaji}, \citenamefont {Monchesky}, \citenamefont {Gordon},\
  and\ \citenamefont {Crozier}}]{Xiao:2009jpcs}%
  \BibitemOpen
  \bibfield  {author} {\bibinfo {author} {\bibfnamefont {Q.~F.}\ \bibnamefont
  {Xiao}}, \bibinfo {author} {\bibfnamefont {S.}~\bibnamefont {Kahwaji}},
  \bibinfo {author} {\bibfnamefont {T.~L.}\ \bibnamefont {Monchesky}}, \bibinfo
  {author} {\bibfnamefont {R.~A.}\ \bibnamefont {Gordon}}, \ and\ \bibinfo
  {author} {\bibfnamefont {E.~D.}\ \bibnamefont {Crozier}},\ }\href@noop {}
  {\bibfield  {journal} {\bibinfo  {journal} {J. Phys.: Conf. Ser.}\ }\textbf
  {\bibinfo {volume} {190}},\ \bibinfo {pages} {012101} (\bibinfo {year}
  {2009})}\BibitemShut {NoStop}%
\bibitem [{\citenamefont {Kahwaji}\ \emph {et~al.}(2012)\citenamefont
  {Kahwaji}, \citenamefont {Gordon}, \citenamefont {Crozier},\ and\
  \citenamefont {Monchesky}}]{Kahwaji:2012prb}%
  \BibitemOpen
  \bibfield  {author} {\bibinfo {author} {\bibfnamefont {S.}~\bibnamefont
  {Kahwaji}}, \bibinfo {author} {\bibfnamefont {R.~A.}\ \bibnamefont {Gordon}},
  \bibinfo {author} {\bibfnamefont {E.~D.}\ \bibnamefont {Crozier}}, \ and\
  \bibinfo {author} {\bibfnamefont {T.~L.}\ \bibnamefont {Monchesky}},\ }\href
  {\doibase 10.1103/PhysRevB.85.014405} {\bibfield  {journal} {\bibinfo
  {journal} {Phys. Rev. B}\ }\textbf {\bibinfo {volume} {85}},\ \bibinfo
  {pages} {014405} (\bibinfo {year} {2012})}\BibitemShut {NoStop}%
\bibitem [{\citenamefont {Hortamani}\ \emph {et~al.}(2006)\citenamefont
  {Hortamani}, \citenamefont {Wu}, \citenamefont {Kratzer},\ and\ \citenamefont
  {Scheffler}}]{Hortamani:2006prb}%
  \BibitemOpen
  \bibfield  {author} {\bibinfo {author} {\bibfnamefont {M.}~\bibnamefont
  {Hortamani}}, \bibinfo {author} {\bibfnamefont {H.}~\bibnamefont {Wu}},
  \bibinfo {author} {\bibfnamefont {P.}~\bibnamefont {Kratzer}}, \ and\
  \bibinfo {author} {\bibfnamefont {M.}~\bibnamefont {Scheffler}},\ }\href
  {\doibase 10.1103/PhysRevB.74.205305} {\bibfield  {journal} {\bibinfo
  {journal} {Phys. Rev. B}\ }\textbf {\bibinfo {volume} {74}},\ \bibinfo
  {pages} {205305} (\bibinfo {year} {2006})}\BibitemShut {NoStop}%
\bibitem [{\citenamefont {Zhao}\ \emph {et~al.}(1992)\citenamefont {Zhao},
  \citenamefont {Jia},\ and\ \citenamefont {Yang}}]{Zhao1992}%
  \BibitemOpen
  \bibfield  {author} {\bibinfo {author} {\bibfnamefont {R.}~\bibnamefont
  {Zhao}}, \bibinfo {author} {\bibfnamefont {J.}~\bibnamefont {Jia}}, \ and\
  \bibinfo {author} {\bibfnamefont {W.}~\bibnamefont {Yang}},\ }\href {\doibase
  10.1016/0039-6028(92)90515-8} {\bibfield  {journal} {\bibinfo  {journal}
  {Surf. Sci.}\ }\textbf {\bibinfo {volume} {274}},\ \bibinfo {pages} {L519 }
  (\bibinfo {year} {1992})}\BibitemShut {NoStop}%
\bibitem [{\citenamefont {Li}\ \emph {et~al.}(1994)\citenamefont {Li},
  \citenamefont {Koziol}, \citenamefont {Wurm}, \citenamefont {Hong},
  \citenamefont {Bauer},\ and\ \citenamefont {Tsong}}]{Li:1994PRB}%
  \BibitemOpen
  \bibfield  {author} {\bibinfo {author} {\bibfnamefont {L.}~\bibnamefont
  {Li}}, \bibinfo {author} {\bibfnamefont {C.}~\bibnamefont {Koziol}}, \bibinfo
  {author} {\bibfnamefont {K.}~\bibnamefont {Wurm}}, \bibinfo {author}
  {\bibfnamefont {Y.}~\bibnamefont {Hong}}, \bibinfo {author} {\bibfnamefont
  {E.}~\bibnamefont {Bauer}}, \ and\ \bibinfo {author} {\bibfnamefont
  {I.~S.~T.}\ \bibnamefont {Tsong}},\ }\href {\doibase
  10.1103/PhysRevB.50.10834} {\bibfield  {journal} {\bibinfo  {journal} {Phys.
  Rev. B}\ }\textbf {\bibinfo {volume} {50}},\ \bibinfo {pages} {10834}
  (\bibinfo {year} {1994})}\BibitemShut {NoStop}%
\bibitem [{\citenamefont {Andersen}\ \emph {et~al.}(1971)\citenamefont
  {Andersen}, \citenamefont {Andreasen}, \citenamefont {Davies},\ and\
  \citenamefont {Uggerh⊘j}}]{Andersen:1971RE}%
  \BibitemOpen
  \bibfield  {author} {\bibinfo {author} {\bibfnamefont {J.~U.}\ \bibnamefont
  {Andersen}}, \bibinfo {author} {\bibfnamefont {O.}~\bibnamefont {Andreasen}},
  \bibinfo {author} {\bibfnamefont {J.~A.}\ \bibnamefont {Davies}}, \ and\
  \bibinfo {author} {\bibfnamefont {E.}~\bibnamefont {Uggerh⊘j}},\ }\href
  {\doibase 10.1080/00337577108232561} {\bibfield  {journal} {\bibinfo
  {journal} {Radiation Effects}\ }\textbf {\bibinfo {volume} {7}},\ \bibinfo
  {pages} {25} (\bibinfo {year} {1971})}\BibitemShut {NoStop}%
\bibitem [{\citenamefont {Heald}\ \emph {et~al.}(1999)\citenamefont {Heald},
  \citenamefont {Brewe}, \citenamefont {Stern}, \citenamefont {Kim},
  \citenamefont {Brown}, \citenamefont {Jiang}, \citenamefont {Crozier},\ and\
  \citenamefont {Gordon}}]{Heald1999JSR}%
  \BibitemOpen
  \bibfield  {author} {\bibinfo {author} {\bibfnamefont {S.~M.}\ \bibnamefont
  {Heald}}, \bibinfo {author} {\bibfnamefont {D.~L.}\ \bibnamefont {Brewe}},
  \bibinfo {author} {\bibfnamefont {E.~A.}\ \bibnamefont {Stern}}, \bibinfo
  {author} {\bibfnamefont {K.~H.}\ \bibnamefont {Kim}}, \bibinfo {author}
  {\bibfnamefont {F.~C.}\ \bibnamefont {Brown}}, \bibinfo {author}
  {\bibfnamefont {D.~T.}\ \bibnamefont {Jiang}}, \bibinfo {author}
  {\bibfnamefont {E.~D.}\ \bibnamefont {Crozier}}, \ and\ \bibinfo {author}
  {\bibfnamefont {R.~A.}\ \bibnamefont {Gordon}},\ }\href {\doibase
  10.1107/S090904959801677X} {\bibfield  {journal} {\bibinfo  {journal}
  {Journal of Synchrotron Radiation}\ }\textbf {\bibinfo {volume} {6}},\
  \bibinfo {pages} {347} (\bibinfo {year} {1999})}\BibitemShut {NoStop}%
\bibitem [{\citenamefont {Bauchspiess}(1991)}]{Rudolf}%
  \BibitemOpen
  \bibfield  {author} {\bibinfo {author} {\bibfnamefont {K.~R.}\ \bibnamefont
  {Bauchspiess}},\ }\emph {\bibinfo {title} {A Study of the Pressure-induced
  Mixed-valence Transition in SmSe and SmS by X-ray Absorption Spectroscopy}},\
  \href@noop {} {Ph.D. thesis},\ \bibinfo  {school} {Simon Fraser University}
  (\bibinfo {year} {1991})\BibitemShut {NoStop}%
\bibitem [{\citenamefont {Bunker}(2010)}]{Bunker}%
  \BibitemOpen
  \bibfield  {author} {\bibinfo {author} {\bibfnamefont {G.}~\bibnamefont
  {Bunker}},\ }\href@noop {} {\emph {\bibinfo {title} {Introduction to XAFS: a
  practical guide to X-ray absorption fine structure spectroscopy}}}\ (\bibinfo
   {publisher} {Cambridge University Press},\ \bibinfo {address} {Cambridge,
  UK},\ \bibinfo {year} {2010})\BibitemShut {NoStop}%
\bibitem [{\citenamefont {Robertson}\ \emph {et~al.}(2006)\citenamefont
  {Robertson}, \citenamefont {Burns},\ and\ \citenamefont
  {Morrison}}]{Robertson:2006ws}%
  \BibitemOpen
  \bibfield  {author} {\bibinfo {author} {\bibfnamefont {M.~D.}\ \bibnamefont
  {Robertson}}, \bibinfo {author} {\bibfnamefont {T.}~\bibnamefont {Burns}}, \
  and\ \bibinfo {author} {\bibfnamefont {T.}~\bibnamefont {Morrison}},\
  }\href@noop {} {\bibfield  {journal} {\bibinfo  {journal} {Micros. Soc. Can.
  Bull.}\ }\textbf {\bibinfo {volume} {34}},\ \bibinfo {pages} {19} (\bibinfo
  {year} {2006})}\BibitemShut {NoStop}%
\bibitem [{\citenamefont {Lian}\ and\ \citenamefont
  {Chen}(1986)}]{Lian:1986apl}%
  \BibitemOpen
  \bibfield  {author} {\bibinfo {author} {\bibfnamefont {Y.~C.}\ \bibnamefont
  {Lian}}\ and\ \bibinfo {author} {\bibfnamefont {L.~J.}\ \bibnamefont
  {Chen}},\ }\href {http://link.aip.org/link/?APL/48/359/1} {\bibfield
  {journal} {\bibinfo  {journal} {Appl. Phys. Lett.}\ }\textbf {\bibinfo
  {volume} {48}},\ \bibinfo {pages} {359} (\bibinfo {year} {1986})}\BibitemShut
  {NoStop}%
\bibitem [{\citenamefont {Lippitz}\ \emph {et~al.}(2005)\citenamefont
  {Lippitz}, \citenamefont {Paggel},\ and\ \citenamefont
  {Fumagalli}}]{Lippitz}%
  \BibitemOpen
  \bibfield  {author} {\bibinfo {author} {\bibfnamefont {H.}~\bibnamefont
  {Lippitz}}, \bibinfo {author} {\bibfnamefont {J.~J.}\ \bibnamefont {Paggel}},
  \ and\ \bibinfo {author} {\bibfnamefont {P.}~\bibnamefont {Fumagalli}},\
  }\href@noop {} {\bibfield  {journal} {\bibinfo  {journal} {Surface Science}\
  }\textbf {\bibinfo {volume} {575}},\ \bibinfo {pages} {307} (\bibinfo {year}
  {2005})}\BibitemShut {NoStop}%
\bibitem [{\citenamefont {Hortamani}\ \emph {et~al.}(2008)\citenamefont
  {Hortamani}, \citenamefont {Sandratskii}, \citenamefont {Kratzer},
  \citenamefont {Mertig},\ and\ \citenamefont {Scheffler}}]{HortamaniPRB78}%
  \BibitemOpen
  \bibfield  {author} {\bibinfo {author} {\bibfnamefont {M.}~\bibnamefont
  {Hortamani}}, \bibinfo {author} {\bibfnamefont {L.}~\bibnamefont
  {Sandratskii}}, \bibinfo {author} {\bibfnamefont {P.}~\bibnamefont
  {Kratzer}}, \bibinfo {author} {\bibfnamefont {I.}~\bibnamefont {Mertig}}, \
  and\ \bibinfo {author} {\bibfnamefont {M.}~\bibnamefont {Scheffler}},\
  }\href@noop {} {\bibfield  {journal} {\bibinfo  {journal} {Phys.Rev.B}\
  }\textbf {\bibinfo {volume} {78}},\ \bibinfo {pages} {104402} (\bibinfo
  {year} {2008})}\BibitemShut {NoStop}%
\bibitem [{\citenamefont {Zhou}\ \emph {et~al.}(2009)\citenamefont {Zhou},
  \citenamefont {Shalimov}, \citenamefont {Potzger}, \citenamefont {Helm},
  \citenamefont {Fassbender},\ and\ \citenamefont {Schmidt}}]{Zhou:2009prb}%
  \BibitemOpen
  \bibfield  {author} {\bibinfo {author} {\bibfnamefont {S.}~\bibnamefont
  {Zhou}}, \bibinfo {author} {\bibfnamefont {A.}~\bibnamefont {Shalimov}},
  \bibinfo {author} {\bibfnamefont {K.}~\bibnamefont {Potzger}}, \bibinfo
  {author} {\bibfnamefont {M.}~\bibnamefont {Helm}}, \bibinfo {author}
  {\bibfnamefont {J.}~\bibnamefont {Fassbender}}, \ and\ \bibinfo {author}
  {\bibfnamefont {H.}~\bibnamefont {Schmidt}},\ }\href {\doibase
  10.1103/PhysRevB.80.174423} {\bibfield  {journal} {\bibinfo  {journal} {Phys.
  Rev. B}\ }\textbf {\bibinfo {volume} {80}},\ \bibinfo {pages} {174423}
  (\bibinfo {year} {2009})}\BibitemShut {NoStop}%
\bibitem [{\citenamefont {Karhu}\ \emph {et~al.}(2010)\citenamefont {Karhu},
  \citenamefont {Kahwaji}, \citenamefont {Monchesky}, \citenamefont {Parsons},
  \citenamefont {Robertson},\ and\ \citenamefont {Maunders}}]{KarhuPRB82}%
  \BibitemOpen
  \bibfield  {author} {\bibinfo {author} {\bibfnamefont {E.}~\bibnamefont
  {Karhu}}, \bibinfo {author} {\bibfnamefont {S.}~\bibnamefont {Kahwaji}},
  \bibinfo {author} {\bibfnamefont {T.~L.}\ \bibnamefont {Monchesky}}, \bibinfo
  {author} {\bibfnamefont {C.}~\bibnamefont {Parsons}}, \bibinfo {author}
  {\bibfnamefont {M.~D.}\ \bibnamefont {Robertson}}, \ and\ \bibinfo {author}
  {\bibfnamefont {C.}~\bibnamefont {Maunders}},\ }\href@noop {} {\bibfield
  {journal} {\bibinfo  {journal} {Phys.Rev.B}\ }\textbf {\bibinfo {volume}
  {82}},\ \bibinfo {pages} {184417} (\bibinfo {year} {2010})}\BibitemShut
  {NoStop}%
\bibitem [{\citenamefont {Wolska}\ \emph {et~al.}(2007)\citenamefont {Wolska},
  \citenamefont {Lawniczak-Jablonska}, \citenamefont {Klepka}, \citenamefont
  {Walczak},\ and\ \citenamefont {Misiuk}}]{WolskaPRB75}%
  \BibitemOpen
  \bibfield  {author} {\bibinfo {author} {\bibfnamefont {A.}~\bibnamefont
  {Wolska}}, \bibinfo {author} {\bibfnamefont {K.}~\bibnamefont
  {Lawniczak-Jablonska}}, \bibinfo {author} {\bibfnamefont {M.}~\bibnamefont
  {Klepka}}, \bibinfo {author} {\bibfnamefont {M.~S.}\ \bibnamefont {Walczak}},
  \ and\ \bibinfo {author} {\bibfnamefont {A.}~\bibnamefont {Misiuk}},\ }\href
  {\doibase 10.1103/PhysRevB.75.113201} {\bibfield  {journal} {\bibinfo
  {journal} {Phys. Rev. B}\ }\textbf {\bibinfo {volume} {75}},\ \bibinfo
  {pages} {113201} (\bibinfo {year} {2007})}\BibitemShut {NoStop}%
\bibitem [{\citenamefont {Li}\ \emph {et~al.}(2007)\citenamefont {Li},
  \citenamefont {Zeng}, \citenamefont {van Benthem}, \citenamefont {Chisholm},
  \citenamefont {Shen}, \citenamefont {Nageswara~Rao}, \citenamefont {Dixit},
  \citenamefont {Feldman}, \citenamefont {Petukhov}, \citenamefont {Foygel},\
  and\ \citenamefont {Weitering}}]{Li:2007PRB}%
  \BibitemOpen
  \bibfield  {author} {\bibinfo {author} {\bibfnamefont {A.~P.}\ \bibnamefont
  {Li}}, \bibinfo {author} {\bibfnamefont {C.}~\bibnamefont {Zeng}}, \bibinfo
  {author} {\bibfnamefont {K.}~\bibnamefont {van Benthem}}, \bibinfo {author}
  {\bibfnamefont {M.~F.}\ \bibnamefont {Chisholm}}, \bibinfo {author}
  {\bibfnamefont {J.}~\bibnamefont {Shen}}, \bibinfo {author} {\bibfnamefont
  {S.~V.~S.}\ \bibnamefont {Nageswara~Rao}}, \bibinfo {author} {\bibfnamefont
  {S.~K.}\ \bibnamefont {Dixit}}, \bibinfo {author} {\bibfnamefont {L.~C.}\
  \bibnamefont {Feldman}}, \bibinfo {author} {\bibfnamefont {A.~G.}\
  \bibnamefont {Petukhov}}, \bibinfo {author} {\bibfnamefont {M.}~\bibnamefont
  {Foygel}}, \ and\ \bibinfo {author} {\bibfnamefont {H.~H.}\ \bibnamefont
  {Weitering}},\ }\href {\doibase 10.1103/PhysRevB.75.201201} {\bibfield
  {journal} {\bibinfo  {journal} {Phys. Rev. B}\ }\textbf {\bibinfo {volume}
  {75}},\ \bibinfo {pages} {201201} (\bibinfo {year} {2007})}\BibitemShut
  {NoStop}%
\bibitem [{\citenamefont {Karhu}\ \emph {et~al.}(2012)\citenamefont {Karhu},
  \citenamefont {R\"o\ss{}ler}, \citenamefont {Bogdanov}, \citenamefont
  {Kahwaji}, \citenamefont {Kirby}, \citenamefont {Fritzsche}, \citenamefont
  {Robertson}, \citenamefont {Majkrzak},\ and\ \citenamefont
  {Monchesky}}]{Karhu:2012prb}%
  \BibitemOpen
  \bibfield  {author} {\bibinfo {author} {\bibfnamefont {E.~A.}\ \bibnamefont
  {Karhu}}, \bibinfo {author} {\bibfnamefont {U.~K.}\ \bibnamefont
  {R\"o\ss{}ler}}, \bibinfo {author} {\bibfnamefont {A.~N.}\ \bibnamefont
  {Bogdanov}}, \bibinfo {author} {\bibfnamefont {S.}~\bibnamefont {Kahwaji}},
  \bibinfo {author} {\bibfnamefont {B.~J.}\ \bibnamefont {Kirby}}, \bibinfo
  {author} {\bibfnamefont {H.}~\bibnamefont {Fritzsche}}, \bibinfo {author}
  {\bibfnamefont {M.~D.}\ \bibnamefont {Robertson}}, \bibinfo {author}
  {\bibfnamefont {C.~F.}\ \bibnamefont {Majkrzak}}, \ and\ \bibinfo {author}
  {\bibfnamefont {T.~L.}\ \bibnamefont {Monchesky}},\ }\href {\doibase
  10.1103/PhysRevB.85.094429} {\bibfield  {journal} {\bibinfo  {journal} {Phys.
  Rev. B}\ }\textbf {\bibinfo {volume} {85}},\ \bibinfo {pages} {094429}
  (\bibinfo {year} {2012})}\BibitemShut {NoStop}%
\bibitem [{\citenamefont {Zeng}\ \emph
  {et~al.}(2008{\natexlab{a}})\citenamefont {Zeng}, \citenamefont {Zhang},
  \citenamefont {van Benthem}, \citenamefont {Chisholm},\ and\ \citenamefont
  {Weitering}}]{Zeng:2008PRL}%
  \BibitemOpen
  \bibfield  {author} {\bibinfo {author} {\bibfnamefont {C.}~\bibnamefont
  {Zeng}}, \bibinfo {author} {\bibfnamefont {Z.}~\bibnamefont {Zhang}},
  \bibinfo {author} {\bibfnamefont {K.}~\bibnamefont {van Benthem}}, \bibinfo
  {author} {\bibfnamefont {M.~F.}\ \bibnamefont {Chisholm}}, \ and\ \bibinfo
  {author} {\bibfnamefont {H.~H.}\ \bibnamefont {Weitering}},\ }\href {\doibase
  10.1103/PhysRevLett.100.066101} {\bibfield  {journal} {\bibinfo  {journal}
  {Phys. Rev. Lett.}\ }\textbf {\bibinfo {volume} {100}},\ \bibinfo {pages}
  {066101} (\bibinfo {year} {2008}{\natexlab{a}})}\BibitemShut {NoStop}%
\bibitem [{\citenamefont {Zeng}\ \emph
  {et~al.}(2008{\natexlab{b}})\citenamefont {Zeng}, \citenamefont {Helgren},
  \citenamefont {Rahimi}, \citenamefont {Hellman}, \citenamefont {Islam},
  \citenamefont {Wilkens}, \citenamefont {Culbertson},\ and\ \citenamefont
  {Smith}}]{Zeng:2008prb}%
  \BibitemOpen
  \bibfield  {author} {\bibinfo {author} {\bibfnamefont {L.}~\bibnamefont
  {Zeng}}, \bibinfo {author} {\bibfnamefont {E.}~\bibnamefont {Helgren}},
  \bibinfo {author} {\bibfnamefont {M.}~\bibnamefont {Rahimi}}, \bibinfo
  {author} {\bibfnamefont {F.}~\bibnamefont {Hellman}}, \bibinfo {author}
  {\bibfnamefont {R.}~\bibnamefont {Islam}}, \bibinfo {author} {\bibfnamefont
  {B.~J.}\ \bibnamefont {Wilkens}}, \bibinfo {author} {\bibfnamefont {R.~J.}\
  \bibnamefont {Culbertson}}, \ and\ \bibinfo {author} {\bibfnamefont {D.~J.}\
  \bibnamefont {Smith}},\ }\href {\doibase 10.1103/PhysRevB.77.073306}
  {\bibfield  {journal} {\bibinfo  {journal} {Phys. Rev. B}\ }\textbf {\bibinfo
  {volume} {77}},\ \bibinfo {pages} {073306} (\bibinfo {year}
  {2008}{\natexlab{b}})}\BibitemShut {NoStop}%
\bibitem [{\citenamefont {Kahwaji}\ \emph {et~al.}()\citenamefont {Kahwaji},
  \citenamefont {Robertson},\ and\ \citenamefont {Monchesky}}]{Kahwaji:unpub}%
  \BibitemOpen
  \bibfield  {author} {\bibinfo {author} {\bibfnamefont {S.}~\bibnamefont
  {Kahwaji}}, \bibinfo {author} {\bibfnamefont {M.}~\bibnamefont {Robertson}},
  \ and\ \bibinfo {author} {\bibfnamefont {T.~L.}\ \bibnamefont {Monchesky}},\
  }\href@noop {} {\bibinfo  {journal} {unpublished}\ }\BibitemShut {NoStop}%
\bibitem [{\citenamefont {Froyen}\ \emph {et~al.}(1989)\citenamefont {Froyen},
  \citenamefont {Wood},\ and\ \citenamefont {Zunger}}]{froyen}%
  \BibitemOpen
\bibfield  {journal} {  }\bibfield  {author} {\bibinfo {author} {\bibfnamefont
  {S.}~\bibnamefont {Froyen}}, \bibinfo {author} {\bibfnamefont {D.~M.}\
  \bibnamefont {Wood}}, \ and\ \bibinfo {author} {\bibfnamefont
  {A.}~\bibnamefont {Zunger}},\ }\href {\doibase 10.1063/1.101100} {\bibfield
  {journal} {\bibinfo  {journal} {Appl. Phys. Lett.}\ }\textbf {\bibinfo
  {volume} {54}},\ \bibinfo {pages} {2435} (\bibinfo {year}
  {1989})}\BibitemShut {NoStop}%
\bibitem [{\citenamefont {Gonz\'alez-M\'endez}\ and\ \citenamefont
  {Takeuchi}(1998)}]{theosurf}%
  \BibitemOpen
  \bibfield  {author} {\bibinfo {author} {\bibfnamefont {M.~E.}\ \bibnamefont
  {Gonz\'alez-M\'endez}}\ and\ \bibinfo {author} {\bibfnamefont
  {N.}~\bibnamefont {Takeuchi}},\ }\href {\doibase 10.1103/PhysRevB.58.16172}
  {\bibfield  {journal} {\bibinfo  {journal} {Phys. Rev. B}\ }\textbf {\bibinfo
  {volume} {58}},\ \bibinfo {pages} {16172} (\bibinfo {year}
  {1998})}\BibitemShut {NoStop}%
\bibitem [{\citenamefont {Tono}\ \emph {et~al.}(2000)\citenamefont {Tono},
  \citenamefont {Yeom}, \citenamefont {Matsuda},\ and\ \citenamefont
  {Ohta}}]{expsurf}%
  \BibitemOpen
  \bibfield  {author} {\bibinfo {author} {\bibfnamefont {K.}~\bibnamefont
  {Tono}}, \bibinfo {author} {\bibfnamefont {H.~W.}\ \bibnamefont {Yeom}},
  \bibinfo {author} {\bibfnamefont {I.}~\bibnamefont {Matsuda}}, \ and\
  \bibinfo {author} {\bibfnamefont {T.}~\bibnamefont {Ohta}},\ }\href {\doibase
  10.1103/PhysRevB.61.15866} {\bibfield  {journal} {\bibinfo  {journal} {Phys.
  Rev. B}\ }\textbf {\bibinfo {volume} {61}},\ \bibinfo {pages} {15866}
  (\bibinfo {year} {2000})}\BibitemShut {NoStop}%
\bibitem [{\citenamefont {Kresse}\ and\ \citenamefont {Hafner}(1994)}]{vasp1}%
  \BibitemOpen
  \bibfield  {author} {\bibinfo {author} {\bibfnamefont {G.}~\bibnamefont
  {Kresse}}\ and\ \bibinfo {author} {\bibfnamefont {J.}~\bibnamefont
  {Hafner}},\ }\href {\doibase 10.1103/PhysRevB.49.14251} {\bibfield  {journal}
  {\bibinfo  {journal} {Phys. Rev. B}\ }\textbf {\bibinfo {volume} {49}},\
  \bibinfo {pages} {14251} (\bibinfo {year} {1994})}\BibitemShut {NoStop}%
\bibitem [{\citenamefont {Kresse}\ and\ \citenamefont
  {Furthm\"{u}ller}(1996)}]{vasp2}%
  \BibitemOpen
  \bibfield  {author} {\bibinfo {author} {\bibfnamefont {G.}~\bibnamefont
  {Kresse}}\ and\ \bibinfo {author} {\bibfnamefont {J.}~\bibnamefont
  {Furthm\"{u}ller}},\ }\href {\doibase 10.1016/0927-0256(96)00008-0}
  {\bibfield  {journal} {\bibinfo  {journal} {Comput. Matter. Sci.}\ }\textbf
  {\bibinfo {volume} {6}},\ \bibinfo {pages} {15 } (\bibinfo {year}
  {1996})}\BibitemShut {NoStop}%
\bibitem [{\citenamefont {Bl\"ochl}(1994)}]{pbe1}%
  \BibitemOpen
  \bibfield  {author} {\bibinfo {author} {\bibfnamefont {P.~E.}\ \bibnamefont
  {Bl\"ochl}},\ }\href {\doibase 10.1103/PhysRevB.50.17953} {\bibfield
  {journal} {\bibinfo  {journal} {Phys. Rev. B}\ }\textbf {\bibinfo {volume}
  {50}},\ \bibinfo {pages} {17953} (\bibinfo {year} {1994})}\BibitemShut
  {NoStop}%
\bibitem [{\citenamefont {Kresse}\ and\ \citenamefont {Joubert}(1999)}]{pbe2}%
  \BibitemOpen
  \bibfield  {author} {\bibinfo {author} {\bibfnamefont {G.}~\bibnamefont
  {Kresse}}\ and\ \bibinfo {author} {\bibfnamefont {D.}~\bibnamefont
  {Joubert}},\ }\href {\doibase 10.1103/PhysRevB.59.1758} {\bibfield  {journal}
  {\bibinfo  {journal} {Phys. Rev. B}\ }\textbf {\bibinfo {volume} {59}},\
  \bibinfo {pages} {1758} (\bibinfo {year} {1999})}\BibitemShut {NoStop}%
\end{thebibliography}
\end{document}